\documentclass[twocolumn,astrosymb,tighten,trackchanges]{aastex631}

\usepackage{color}
\usepackage{colortbl}
\definecolor{YG}{RGB}{115,80,185}

\graphicspath{{./}{figures/}}

\begin{document}

\title{Late-time Hubble Space Telescope Observations of AT\,2018cow. I. Further Constraints on the Fading Prompt Emission and Thermal Properties 50--60 Days Post-discovery}

\correspondingauthor{Yuyang Chen}
\email{yuyangf.chen@mail.utoronto.ca}

\author[0000-0002-8804-3501]{Yuyang Chen}
\affiliation{David A. Dunlap Department of Astronomy and Astrophysics, University of Toronto, 50 St. George Street, Toronto M5S 3H4, Canada}
\affiliation{Dunlap Institute for Astronomy and Astrophysics, University of Toronto, 50 St. George Street, Toronto M5S 3H4, Canada}

\author[0000-0001-7081-0082]{Maria R. Drout}
\affiliation{David A. Dunlap Department of Astronomy and Astrophysics, University of Toronto, 50 St. George Street, Toronto M5S 3H4, Canada}
\affiliation{The Observatories of the Carnegie Institution for Science, 813 Santa Barbara St., Pasadena, CA 91101, USA}

\author[0000-0001-6806-0673]{Anthony L. Piro}
\affiliation{The Observatories of the Carnegie Institution for Science, 813 Santa Barbara St., Pasadena, CA 91101, USA}

\author[0000-0002-5740-7747]{Charles~D.~Kilpatrick}
\affiliation{Department of Physics and Astronomy, Northwestern University, Evanston, IL 60208, USA}
\affiliation{Center for Interdisciplinary Exploration and Research in Astrophysics (CIERA), 1800 Sherman Ave, Evanston, IL 60201, USA}

\author[0000-0002-2445-5275]{Ryan~J.~Foley}
\affiliation{Department of Astronomy and Astrophysics, University of California, Santa Cruz, CA 95064, USA}

\author[0000-0002-7559-315X]{C\'{e}sar Rojas-Bravo}
\affiliation{Department of Astronomy and Astrophysics, University of California, Santa Cruz, CA 95064, USA}

\author[0000-0002-5748-4558]{Kirsty~Taggart}
\affiliation{Department of Astronomy and Astrophysics, University of California, Santa Cruz, CA 95064, USA}

\author{Matthew R. Siebert}
\affiliation{Space Telescope Science Institute, Baltimore, MD 21218, USA}

\author[0000-0002-0629-8931]{M. R. Magee}
\affiliation{Department of Physics, University of Warwick, Gibbet Hill Road, Coventry CV4 7AL, UK}

\begin{abstract}

The exact nature of the luminous fast blue optical transient AT\,2018cow is still debated. In this first of a two-paper series, we present a detailed analysis of three Hubble Space Telescope (HST) observations of AT\,2018cow covering $\sim$50--60 days post-discovery in combination with other observations throughout the first two months, and derive significantly improved constraints of the late thermal properties. By modeling the spectral energy distributions (SEDs), we confirm that the UV-optical emission over 50--60 days was still a smooth blackbody (i.e., optically thick) with a high temperature ($T_{\mathrm{BB}}\sim15000\,\mathrm{K}$) and small radius ($R_{\mathrm{BB}}\lesssim1000\,R_\odot$). Additionally, we report for the first time a break in the bolometric light curve: the thermal luminosity initially declined at a rate of $L_{\mathrm{BB}}\propto t^{-2.40}$, but faded much faster at $t^{-3.06}$ after day 13. Reexamining possible late-time power sources, we disfavor significant contributions from radioactive decay based on the required $^{56}$Ni mass and lack of UV line blanketing in the HST SEDs. We argue that the commonly proposed interaction with circumstellar material may face significant challenges in explaining the late thermal properties, particularly the effects of the optical depth. Alternatively, we find that continuous outflow/wind driven by a central engine can still reasonably explain the combination of a receding photosphere, optically thick and rapidly fading emission, and intermediate-width lines. However, the rapid fading may have further implications on the power output and structure of the system. Our findings may support the hypothesis that AT\,2018cow and other ``Cow-like transients'' are powered mainly by accretion onto a central engine.

\end{abstract}

\section{Introduction} \label{sec:introduction}

The recent advent of high-cadence, wide-field optical surveys has unveiled a new class of peculiar transients, commonly termed the fast blue optical transients (FBOTs; adopted hereafter), rapidly-evolving transients (RETs), or fast evolving luminous transients (FELTs). As the name suggests, the defining characteristics of FBOTs are the fast-evolving light curves (time above half-brightness $\lesssim 15\,\mathrm{days}$) and blue color ($g-r \lesssim 0.0$) at peak \citep[e.g.][]{2014ApJ...794...23D,2016ApJ...819....5T,2016ApJ...819...35A,2018NatAs...2..307R,2018MNRAS.481..894P,2020ApJ...894...27T,2020MNRAS.498.2575W,Ho2021}. 

FBOTs have peak magnitudes that span a wide range ($-15 \gtrsim M_{\mathrm{peak}} \gtrsim -22$) and can have vastly varying light curves and spectra, demonstrating diversity within the class itself. As a whole, FBOTs are not intrinsically rare, with an inferred volumetric rate of $\sim 1-10\%$ of the core-collapse supernova (CCSN) rate \citep[][although this depends on the exact definition of an FBOT]{2014ApJ...794...23D,2018MNRAS.481..894P}. However, their extreme timescales pose a challenge to conventional supernova (SN) models that rely on radioactive decay and hydrogen recombination as the primary energy source. 

As a result, a variety of alternative scenarios have been proposed to explain the characteristics of FBOTs, some of which involve interactions with circumstellar material \citep[CSM;][]{2010ApJ...724.1396O,2016MNRAS.461.3057S,2018NatAs...2..307R,McDowell2018,Kleiser2018CSM,Tolstov2019,Wang2019,Suzuki2020,Wang2020,Karamehmetoglu2021,Maeda2022,Margalit2022a,Margalit2022b,Mor2023,Liu2023,Khatami2023}, energy injection by a central engine such as a neutron star \citep[NS;][]{Yu2015,2017ApJ...850...18H,2017ApJ...851..107W,Wang2022,Liu2022} or a black hole \citep[BH;][]{2015MNRAS.451.2656K,2018NatAs...2..307R,Tsuna2021,Fujibayashi2022}, tidal disruption events \citep[TDEs;][]{2020ApJ...890L..26K,Kremer2021}, electron-capture supernovae \citep[ECSNe;][]{2016MNRAS.461.2155M}, and SNe within extended envelopes \citep[][]{Brooks2017,Kleiser2018HeGiant}. 

A recent population study by \citet{Ho2021} revealed that most FBOTs are located in star-forming galaxies and are spectroscopically similar to established SNe types (e.g. Type IIb and Ibn). The authors therefore suggested that most FBOTs are likely extreme variations of known classes of SNe with massive star progenitors, but noted that outliers do exist (e.g., AT\,2020bot located in an elliptical galaxy).

In this vein, recently, a small subset of extremely luminous FBOTs ($M_{\mathrm{peak}}\lesssim -21$) accompanied by bright multi-wavelength emissions\footnote{Bright radio emissions were observed for all luminous FBOTs over the first few hundred days. Bright X-ray emissions were typically observed as well, with the exception of AT\,2018lug which did not have any follow-up X-ray observation (i.e., no confirmed X-ray emission). Also, the X-ray detection of CSS161010 was weaker and relatively less luminous.} have been discovered (sometimes referred to as ``Cow-like transients''). These include AT\,2018cow \citep[``The Cow'';][]{2018ATel11727....1S,2018ApJ...865L...3P}, CSS161010 \citep[][]{Coppejans2020}, AT\,2018lug \citep[``The Koala'';][]{Ho2020}, AT\,2020xnd \citep[``The Camel'';][]{Perley2021Camel,Bright2022Camel,Ho2022Camel}, and AT\,2020mrf \citep[][]{Yao2022}. \citet{Coppejans2020} and \citet{Ho2021} found this sub-population to be rare ($\lesssim 0.1\%$ of the CCSN rate) and possibly be an entirely new class of transients distinct from typical FBOTs and SNe. The exact nature of these luminous FBOTs is still unclear, but studies often associate the observations with extreme CSM or central engine configurations, which could have significant implications on our understanding of stellar death. 

The focus of this study is the first luminous FBOT -- AT\,2018cow -- discovered on 2018-06-16 at 10:35:02 UTC (or MJD 58285.441) by the Asteroid Terrestrial-impact Last Alert System \citep{2018ATel11727....1S,2018ApJ...865L...3P}, located in the star-forming dwarf galaxy CGCG 137-068 ($z=0.0141$). AT\,2018cow rose to a magnitude of $M\sim -22$, exceeding those of some superluminous SNe, in $\lesssim 2$ days. For reference, a typical SLSN can take weeks to rise to a peak \citep[see review by][]{2019NatAs...3..697I}. The early discovery, low redshift, and peculiar natures of AT\,2018cow garnered a great deal of interest and triggered global campaigns of multiwavelength follow-up observations \citep[e.g.,][]{RiveraSandoval2018,Kuin2019,Ho2019,Margutti2019,Perley2019,Huang2019,Bietenholz2020,Xiang2021,Pasham2021}.

Analyses of the observations of AT\,2018cow revealed many remarkable properties, some of which were unprecedented in known transients. Some key properties \citep[mostly described in][]{Perley2019,Margutti2019,Ho2019} are summarized below.
\begin{itemize}
    \item AT\,2018cow displayed UV-bright thermal emission, with a blackbody (BB) luminosity that peaked at $L_{\mathrm{BB}} \sim 4\times10^{44}\,\mathrm{erg}\,\mathrm{s}^{-1}$ and subsequently declined at a rate of $L_{\mathrm{BB}}\propto t^{-\alpha}$ with $\alpha\sim2-2.5$.
    \item The blackbody temperature peaked at $T_{\mathrm{BB}} \sim 30000\,\mathrm{K}$, declined to $\sim 15000\,\mathrm{K}$ after 20 days post-discovery, and stayed roughly constant afterwards.
    \item The blackbody radius started at $R_{\mathrm{BB}} \sim 8\times10^{14}\,\mathrm{cm}$ (or $11000\,R_\odot$) and receded monotonically.
    \item The optical spectra were initially featureless. Some broad absorption features ($v \sim 0.1c$) appeared and disappeared over $\sim4$--15 days post-discovery. Afterwards, intermediate-width ($v \sim 3000-6000\,\mathrm{km}\,\mathrm{s}^{-1}$) He (and some H) emission lines emerged. The intermediate-width lines were initially redshifted, but the line profile became more asymmetric as the peak evolved blueward.
    \item Persistent excess NIR emission was detected at $\lambda \gtrsim 10000\,\mathrm{\AA}$ since discovery.
    \item Bright soft X-ray emission was detected with a peak $L_\mathrm{0.3-10keV} \sim 10^{43}\,\mathrm{erg}\,\mathrm{s}^{-1}$. An initial slow decline ($L_\mathrm{0.3-10keV} \propto t^{-1}$) was followed by a rapid decline ($\propto t^{-4}$) after $\sim 20$ days post-discovery with significant variability. Hard X-ray ($> 15\,\mathrm{keV}$) emission was also detected over the first $\sim15$ days.
    \item Bright mm radio emission was detected, with $\nu L_\nu \sim 10^{40}-10^{41}\,\mathrm{erg}\,\mathrm{s}^{-1}$ ($\nu > 90\,\mathrm{GHz}$) for the first $\sim 50$ days. The emission was consistent with synchrotron emission produced by a blast wave moving at $v \sim 0.1c$ in a dense medium.
\end{itemize}
These properties are inconsistent with simple homologous expansion \citep[][]{Liu2018}, and thus alternative transient models must be considered. Numerous theoretical models were put forth to explain AT\,2018cow, such as a TDE by a non-stellar-mass BH \citep{Perley2019,Kuin2019}, a SN powered by ejecta-CSM interaction \citep[][]{RiveraSandoval2018,Xiang2021,Pellegrino2022}, a failed SN generating wind outflow from a newly formed accretion disk \citep[][]{Quataert2019,Piro2020,Uno2020}, a magnetar-powered SN following the explosion of a star \citep[][]{Margutti2019,Fang2019,Mohan2020} or the collapse of a white dwarf \citep[WD;][]{Lyutikov2019,Yu2019,Lyutikov2022}, a jet from a CCSN \citep[][]{Gottlieb2022}, a delayed merger of a BH-star binary system \citep[][]{Metzger2022}, a pulsational pair-instability SN \citep{Leung2020}, and a common envelope jets supernova \citep[][]{Soker2019,Soker2022,Cohen2023}.

Over time, the TDE hypothesis became less favored due to difficulties explaining the non-detection of linear polarization \citep[][]{Huang2019}, the existence of an intermediate-mass BH (IMBH) or a supermassive BH (SMBH) at the outskirt of the galaxy where the gas velocity is smoothly varying without any signs of a coincident massive host system \citep[][]{Lyman2020}{}{}, and the mass limit of $M_{\mathrm{BH}}<850\,M_\odot$ derived from NICER X-ray quasiperiodic oscillations \citep[QPO;][]{Pasham2021}. The dense CSM around AT\,2018cow would also be difficult to explain unless the BH was already embedded in a gas-rich environment \citep[][]{Margutti2019}{}{}. Host galaxy studies have also found ongoing star-formation activities around AT\,2018cow \citep[][]{Roychowdhury2019,Morokuma-Matsui2019,Lyman2020,Sun2022new}, favoring massive star progenitors. (Although \citealt{Michalowski2019} have argued against associating the star formation with AT\,2018cow.)

Even under the assumption of a massive star progenitor, many models remain viable for explaining AT\,2018cow, and the true nature of this peculiar transient is still very much debated. To disentangle different models of AT\,2018cow, observations at late times can be beneficial for directly probing (i) the immediate surrounding to track star-forming history and associated host cluster or companion star, and (ii) any fading transient emission when the effects of optical depth have eased, useful for distinguishing power sources (e.g., radioactive decay, magnetar spin down, accretion around compact object) that predict different rates of decline. To serve these purposes, the Hubble Space Telescope (HST) has carried out observations of AT\,2018cow at six late-time epochs as of date, with the first three covering $\sim$50--60 days post-discovery and the latest three covering $\sim$2--4 years post-discovery. \citet{Sun2022,Sun2022new} have used the HST images taken years post-discovery to examine the immediate surrounding and reported the discovery of a spatially-coincident underlying source and the possibility that AT\,2018cow was at the foreground relative to the nearby star-forming regions. 

In this study, we present a detailed analysis of the first three HST observations of AT\,2018cow covering 50--60\,days post-discovery combined with other UV-Optical-IR (UVOIR) observations taken throughout the first two months. The high-resolution HST images provide significantly improved constraints on the NUV spectral energy distributions (SEDs) of the fading prompt emission and offer important clues on the evolution of AT\,2018cow. We utilize these late-time observations to place additional constraints on the possible power sources of AT\,2018cow. This is paper I of a two-paper series, while paper II \citep[][]{Chen2023II} focuses on the evolution of the spatially-coincident underlying source in the latest three HST observations.

Note that throughout this two-paper series, we refer to the emission of AT\,2018cow detected over the first two months as \emph{prompt emission}. Although this is synonymous with AT\,2018cow in previous literature, we use this term to distinguish the initial evolution from the spatially-coincident source discovered years post-discovery, which we call separately as the \emph{underlying source}. The motivation for separate terms is outlined in more detail in paper II as it becomes relevant when discussing both emissions. In brief, we make this choice because of the ambiguity in the classification and physical mechanisms associated with the emission observed years later.

The three HST observations and other UVOIR observations of AT\,2018cow used in this study are described in Section \ref{sec:observations}. In Section \ref{sec:promptcow}, we construct and analyze SEDs to derive updated thermal properties of AT\,2018cow. In Section \ref{sec:promptinterpret}, we discuss the updated properties and place additional constraints on the proposed late-time power sources of AT\,2018cow. Finally, we summarize the results and discuss their implications on the nature of AT\,2018cow in Section \ref{sec:conclusion}.

For our analyses, we adopt a luminosity distance of $d_L = 60\,\mathrm{Mpc}$ for AT\,2018cow \citep[][]{Perley2019,Margutti2019}. We assume an $R_{\mathrm{V}} = 3.1$ Milky Way extinction curve with $E_{\mathrm{B}-\mathrm{V}} = 0.07\,\mathrm{mag}$ \citep[][]{2011ApJ...737..103S} and no internal extinction in the host galaxy of AT\,2018cow. Throughout this study, we refer to $t$ as the rest-frame time after the first discovery date of AT\,2018cow, MJD 58285.441 \citep[][]{2018ATel11727....1S,2018ApJ...865L...3P}{}{}.

\begin{figure}
\epsscale{1.13}
\plotone{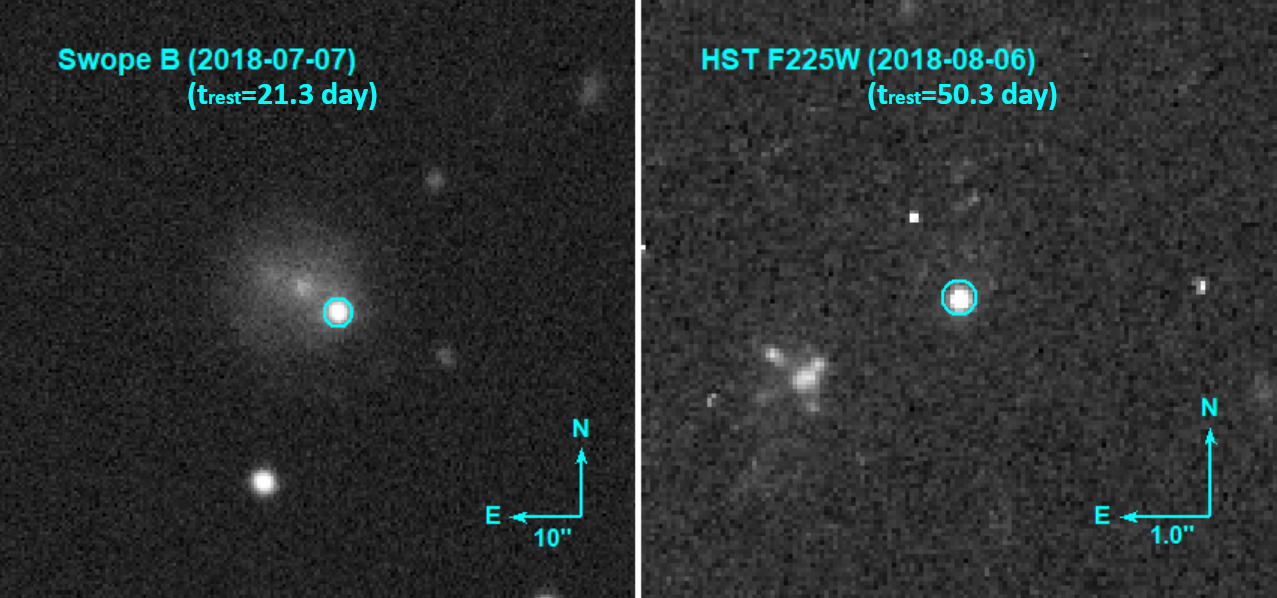}
\caption{Example cutout images of the field of AT\,2018cow taken by Swope (left) and HST (right). The filter and observation date are shown at the top, and a circular aperture is placed at the position of AT\,2018cow. The high-resolution HST image shows AT\,2018cow as an isolated point source not affected by any significant contamination from the diffuse background or nearby sources.
\label{fig:Img}}
\end{figure}

\begin{deluxetable}{cccc}
\tablecaption{HST photometry of AT\,2018cow\label{tab:HSTmag}}
\tablewidth{0pt}
\tablehead{
\colhead{Filter} & \multicolumn{3}{c}{Magnitude (AB)}
\\
\colhead{} & \colhead{$50.3\,\mathrm{d}$} & \colhead{$55.9\,\mathrm{d}$} & \colhead{$60.1\,\mathrm{d}$}
}
\startdata
F218W & $20.89(0.02)$ & $21.29(0.03)$ & $21.54(0.04)$      \\
F225W & $20.68(0.01)$ & $21.08(0.02)$ & $21.30(0.03)$      \\
F275W & $20.37(0.02)$ & $20.72(0.02)$ & $20.97(0.03)$      \\
F336W & $20.12(0.02)$ & $20.69(0.03)$ & $20.81(0.03)$        
\enddata
\tablecomments{1$\sigma$ errors are given inside the brackets.}
\end{deluxetable}

\section{Observations and Data Reduction} \label{sec:observations}

In this section, we describe the observations used throughout this study. While our focus is on new constraints that can be derived from late-time HST observations of AT\,2018cow, our modeling is supplemented by observations from other facilities/wavelengths. In Figure \ref{fig:Img}, we show example images of the containing AT\,2018cow taken by the 1\,m Swope telescope at Las Campanas Obervatory (left) and HST (right) for comparison. In addition to photometric observations described in this section, in Appendix \ref{apdx:spectra}, we also present a series of previously unpublished spectra, including a high-resolution spectrum taken at $t=4.77\,\mathrm{days}$.

\subsection{HST Observations ($t\simeq50-60\,\mathrm{days}$)}\label{sec:HSTprompt}

The HST first observed the prompt emission of AT\,2018cow (PI Foley) with the Wide Field Camera 3 (WFC3) at three late-time epochs (MJD = 58336.4, 58342.1, 58346.4, or $t \simeq 50.3,\,55.9,\,60.1\,\mathrm{days}$). Observations were obtained in four UVIS bands (F218W, F225W, F275W, F336W) spanning $\lambda \simeq 2224\,\mathrm{\AA}-3355\,\mathrm{\AA}$, designed to track the fading UV emission as it fell below the detection limit of \emph{Swift-}UVOT. The HST data can be found on the Mikulski Archive for Space Telescopes (MAST): \dataset[10.17909/5ry6-7r85]{http://dx.doi.org/10.17909/5ry6-7r85}. In the high-resolution HST images, AT\,2018cow is seen as an isolated point source without significant background contamination (e.g., Figure \ref{fig:Img}), allowing precise measurements of its late-time UV flux. 

All HST observations were reduced using the {\tt hst123} pipeline \citep{hst123} as described in \citet{Kilpatrick21}. Each calibrated ({\tt flc}) WFC3/UVIS frame was downloaded from MAST and aligned to a common reference frame using {\tt TweakReg} \citep{drizzlepac}. We then drizzled each individual epoch and band using {\tt drizzlepac} before performing point-spread function (PSF) photometry in the original {\tt flc} frames with {\tt dolphot} \citep{dolphot}. We used standard parameters for {\tt dolphot} as described in the WFC3 guide and {\tt hst123} documentation. As a reference frame for {\tt dolphot}, we used drizzled imaging in WFC3/UVIS F336W obtained at $t\simeq50.3\,\mathrm{days}$. This image was aligned to all {\tt flc} frames both in {\tt TweakReg} and subsequently by alignment methods in {\tt dolphot}, with the rms uncertainty in frame-to-frame alignment ranging from 0.005--0.049\arcsec. Final HST photometry of AT\,2018cow are presented in Table \ref{tab:HSTmag} and shown in Figure \ref{fig:LC} as stars.

We note that \citet{Sun2022} also reported photometry of AT\,2018cow from the first three HST observations, but did not analyze these photometry in their study. We find our photometry to generally be consistent with those reported by \citet{Sun2022}.

\subsection{\emph{Swift}-UVOT Observations ($t\sim3-70\,\mathrm{days}$)}

To provide context for the UV evolution of AT\,2018cow prior to the HST observations, we used data from the Ultra-Violet Optical Telescope \citep[UVOT;][]{2005SSRv..120...95R} onboard the Neil Gehrels Swift Observatory \citep[\emph{Swift};][]{2004ApJ...611.1005G}. \emph{Swift} observations of AT\,2018cow began from $t\simeq3\,\mathrm{days}$ and continued for $\sim$70 days (although later epochs suffered from significant background contamination). The UVOT observations covered all six bands ($uw1$, $uw2$, $um2$, $uvu$, $uvb$, $uvv$), spanning the wavelength range of $\lambda_{\mathrm{eff}} \simeq 2086\,\mathrm{\AA}-5411\,\mathrm{\AA}$. Note that $uvu$, $uvb$, and $uvv$ refer to the UVOT $u$, $b$, and $v$ bands.

While \emph{Swift}-UVOT data of AT\,2018cow have previously been published \citep[e.g.][]{Perley2019,Kuin2019}, in September 2020, a new UVOT sensitivity calibration file was released to account for the loss of sensitivity in the UV and white filters. This retroactively applied to all UVOT observations after 2017, including those of AT\,2018cow, with a possible difference up to 0.3\,mag. For this reason, we performed our own analysis using an updated UVOT \textsc{caldb} (20201008 release).

We downloaded the reduced level 2 sky images of AT\,2018cow from the \emph{Swift} archive. Images were discarded if AT\,2018cow landed on a patch of the detector with reduced sensitivity. Multi-extension images were combined using \texttt{uvotimsum} with default parameters whenever possible. We used \texttt{uvotsource} to perform aperture photometry on AT\,2018cow. The source region was chosen to be a 3\arcsec\ circular aperture, while the background region was chosen to be a 30\arcsec\ circular aperture in an empty sky region away from the host galaxy. Aperture corrections were applied to the photometry using the curve-of-growth method.

\begin{figure*}
\epsscale{1.0}
\plotone{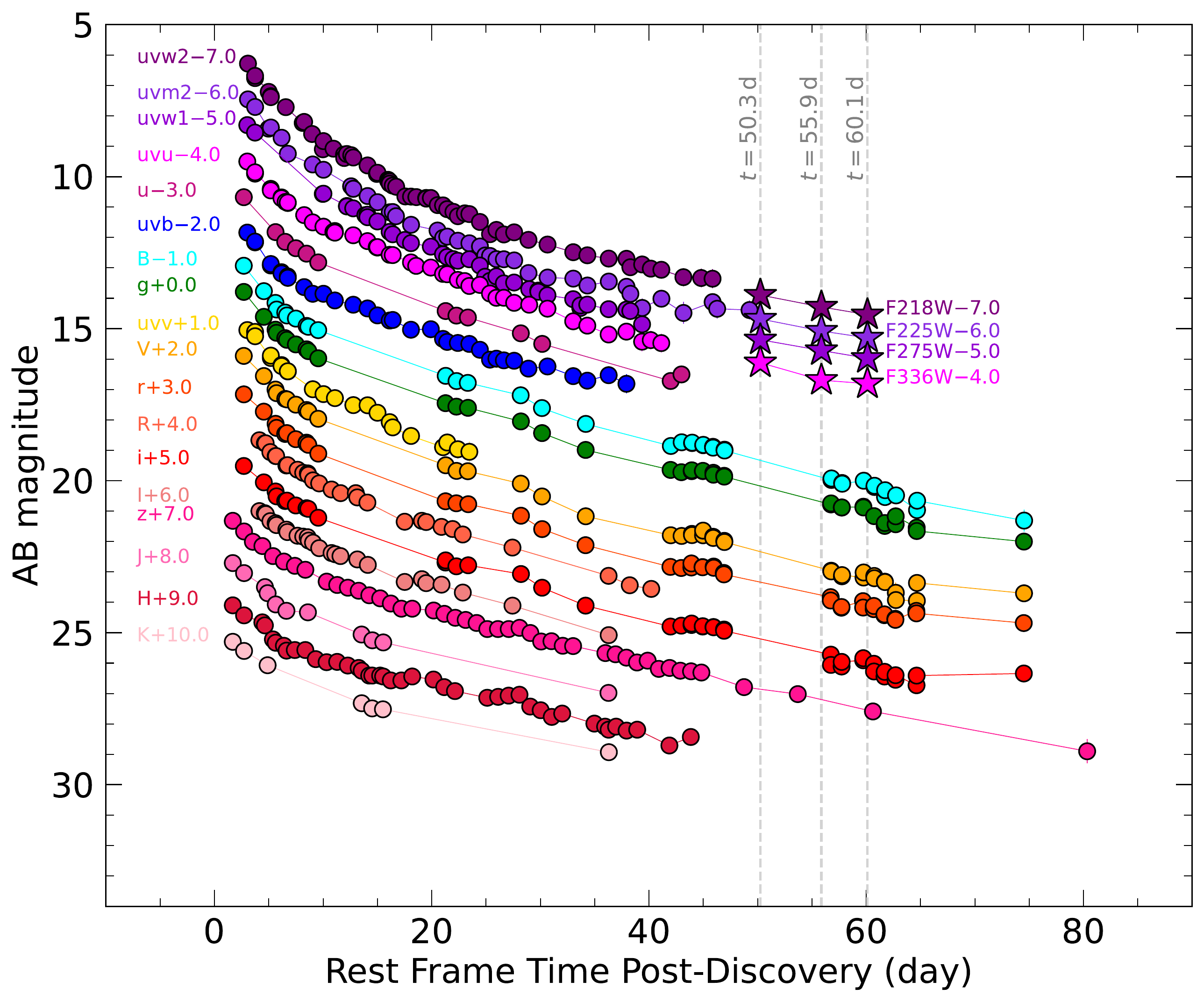}
\caption{The final multi-band UVOIR light curves of AT\,2018cow used in this study. The HST photometry are shown as stars, and the HST epochs are marked with dashed lines. Magnitudes are shifted for visual clarity. UVOT photometry are included only if they are above the $3\sigma$ detection level. The photometry shown is available as Data behind the Figure.
\label{fig:LC}}
\end{figure*}

To account for local background contamination, we applied the same source and background apertures to UVOT images of the host galaxy taken in MJD 58761 (or $t\simeq467\,\mathrm{days}$) and measured the background emission at the location of AT\,2018cow. We then subtracted the measured background flux from the fluxes of AT\,2018cow obtained in the previous step. Based on this analysis, we found that AT\,2018cow was robustly detected ($>3\sigma$) by \emph{Swift} out to $t\sim40-50\,\mathrm{days}$. 

Compared to previously published \emph{Swift}-UVOT photometry in studies of AT\,2018cow \citep[e.g.,][]{Perley2019,Kuin2019}, our photometry are generally very consistent (within error), with differences much less than 0.3\,mag. We note that a more recent paper, \citet{Hinkle2021}, used the updated calibration file and published corrected \emph{Swift}-UVOT photometry of TDE candidates, including AT\,2018cow. However, our photometry (and previously published ones) are inconsistent with the photometry in \citet{Hinkle2021}, with differences up to $\approx0.8$\,mag. Given that these differences are larger than the expected correction, we chose to use photometry derived from our analysis. The $>3\sigma$ \emph{Swift}-UVOT photometry used in this study are shown in Figure \ref{fig:LC}.

\subsection{Swope Observations ($t\sim3-75\,\mathrm{days}$)}

To constrain the SEDs of the prompt emission of AT\,2018cow, particularly at the time of the HST observations, we required optical data at similar epochs. We obtained photometry from the 1\,m Swope Telescope at the Las Campanas Observatory, Chile, which started observing AT\,2018cow at $t\sim3\,\mathrm{days}$ and continued for $\sim$70 days. The observations covered the $uBgVri$ bands, spanning the wavelength range $\lambda_{\mathrm{eff}} \simeq 3608\,\mathrm{\AA} - 7458\,\mathrm{\AA}$. 

Following the reduction procedures described in \citet{kilpatrick2018}, all image processing and optical photometry on the Swope data was performed using {\tt photpipe} \citep[][]{rest2005}, including bias-subtraction, flat-fielding, image
stitching, registration, and photometric calibration. The $BgVri$ photometry were calibrated using standard sources from the Pan-STARRS DR1 catalog \citep{Flewelling2020}, and the $u$-band data were calibrated using SkyMapper u-band standards \citep{onken2019}, both in the same field as AT\,2018cow, and transformed into the Swope natural system \citep{krisciunas2017} with the Supercal method \citep{scolnic2015}.

\begin{figure*}
\epsscale{1.0}
\plotone{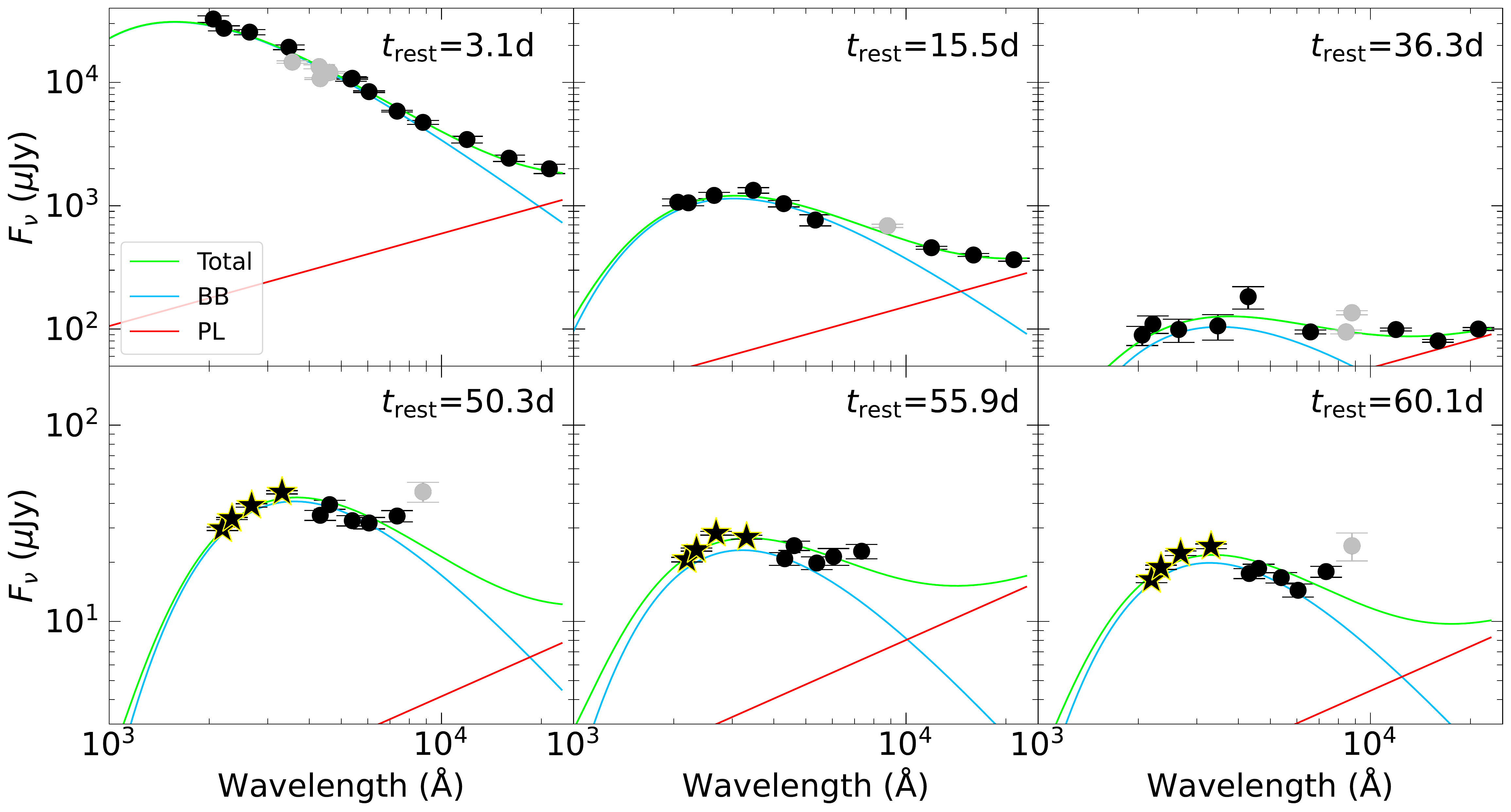}
\caption{The observed (dereddened) SEDs (black) of AT\,2018cow at six sample epochs including the three HST epochs (bottom panels). The HST measurements are marked as stars. The best-fit models (solid lines) are also shown, derived using the \textsc{blackbody + power law} model. Photometry excluded from the fit are shown in gray (see text for reasoning). The UV-optical continuum of AT\,2018cow at $t\simeq50-60\,\mathrm{days}$ is consistent with a blackbody curve peaking at $\lambda \sim 3000\,\mathrm{\AA}$.
\label{fig:SEDfits}}
\end{figure*}

To ensure precise measurements at late times, when background contamination from a nearby star-forming region (Figure \ref{fig:Img}) becomes significant, we obtained deep $uBgVri$ observations between 2019-04-10 and 2021-04-27 with the same telescope and instrument configuration and performed image subtraction using \texttt{hotpants} \citep{becker2015}. Forced photometry was performed on the subtracted images to obtain the final photometry shown in Figure \ref{fig:LC}. We note that our Swope photometry are also generally consistent with previously published photometry in the $uBgVri$ bands from different telescopes \citep[e.g.,][]{Perley2019}.

\subsection{NIR Photometry ($t\sim2-80\,\mathrm{days}$)}

We also make use of the $RIzJHK$ photometry published by \citet{Perley2019}, which span $t\sim2-80\,\mathrm{days}$. These measurements were taken by the 2\,m Liverpool Telescope \citep[][]{2004SPIE.5489..679S}, the 1\,m telescope at the Mount Laguna Observatory \citep[][]{1969PASP...81...74S}, the 1.5\,m telescope at the Cerro Tololo Inter-American Observatory, the 2\,m Himalayan Chandra Telescope, the 0.4\,m telescope at Lulin Observatory, the Palomar 200-inch Hale Telescope, and the MPG/ESO 2.2\,m telescope \citep[][]{2008PASP..120..405G} at La Silla Observatory. For details on image processing, calibration, and host subtraction, see Section 2.2 of \citet{Perley2019}.

\section{Evolution of Prompt UVOIR Emission} \label{sec:promptcow}

Our final multi-band UVOIR light curves from the prompt emission ($t\sim2-80\,\mathrm{days}$) of AT\,2018cow are shown in Figure \ref{fig:LC}.
The addition of the HST measurements provides significantly improved constraints on the UV emission beyond the coverage of \emph{Swift}. In this section, we describe our updated analysis of the prompt UVOIR emission of AT\,2018cow, with an emphasis on the later epochs ($t\sim40-60$ days).

\subsection{Constructing UVOIR SEDs} \label{subsec:constructSED}

To construct the SEDs, we linearly interpolated the light curves to a common set of epochs. We selected a total of 22 epochs, including the three HST epochs, that have sufficient UV-optical spectral coverage. The errors in the interpolated magnitudes were found through Monte Carlo propagation. At each epoch, we only interpolated a given light curve if the nearest measurement was within 1.0 days from the epoch. The only exception was the first HST epoch ($t = 50.3\,\mathrm{days}$) which did not have any measurements in the optical bands within 1.0 days. For this epoch, we interpolated all the optical bands regardless and added 0.03 mag (approximately 3\% of flux) to the uncertainties of the interpolated magnitudes to account for deviation from linear interpolation.

\subsection{SED Shape at $t\simeq50-60\,\mathrm{days}$} \label{subsec:HSTSEDshape}

The precision of the HST photometry allows for the most accurate assessment to date of the UV-optical SED of AT\,2018cow at $t\simeq50-60\,\mathrm{days}$. As shown in the bottom panels of Figure \ref{fig:SEDfits}, the HST photometry trace a smooth curve peaking at $\lambda \sim 3000\,\mathrm{\AA}$ with an exponential decay in the NUV. The shape of the SEDs at these times resembles a blackbody (Section \ref{subsec:SEDmodel}), similar to the previous epochs, suggesting optically thick emission even at these late epochs despite the rapid fading. The smooth continuum in the NUV ($\lambda\sim2000-3000\,\mathrm{\AA}$) also implies a lack of line blanketing from iron peak elements. We highlight the spectral shape here because our results (Section \ref{subsec:updatethermal}) and interpretations (Section \ref{sec:promptinterpret}) are heavily influenced by this property.

\subsection{Modeling of the UVOIR SEDs} \label{subsec:SEDmodel}

We performed forward modeling to characterize the UVOIR SEDs of AT\,2018cow. In summary, we defined a theoretical spectrum to model the emission of the transient, reddened the spectrum according to the Cardelli extinction law \citep{1989ApJ...345..245C} with $R_{\mathrm{V}}=3.1$, and performed synthetic photometry to obtain the model magnitude in each band. Finally, we fit the model SED to the observed SED using the Markov Chain Monte Carlo (MCMC) sampler in the Python package \texttt{emcee} \citep[][]{2013PASP..125..306F}. The best-fit model parameters and uncertainties were derived from the 50th, 15.9th, and 84.1th percentile of the resulting samples.

Regarding the theoretical spectrum, it is well-established that a blackbody can characterize most of the UV-optical emission of AT\,2018cow when the transient was still bright \citep[][]{Perley2019,Margutti2019,Kuin2019}. In addition, a second component was found to be necessary at all times to account for excess IR emission. The main objective of this study is to constrain the late thermal properties of AT\,2018cow using UV measurements from the HST. Therefore, we followed \citet{Perley2019} and adopted a model in the form of \textsc{blackbody + power law} (BB + PL). The thermal properties are extracted directly through the blackbody while the power law describes the excess IR emission phenomenologically. 

In the model \textsc{blackbody + power law}, the blackbody is simply the Planck function $B_\nu(T)$, while the power law takes the form $F_{\nu,\mathrm{PL}} \propto \nu^\alpha$. Following \citet{Perley2019}, the spectral index is fixed at $\alpha = -0.75$ for all epochs. The total flux density from the theoretical spectrum is then given by
\begin{eqnarray}
    f_\nu &=& F_{\nu,\mathrm{BB}}(R_{\mathrm{BB}}, T_{\mathrm{BB}}) + F_{\nu,\mathrm{PL}}(A_{\mathrm{PL}}) \nonumber \\ 
    &=& \frac{\pi R_{\mathrm{BB}}^2}{d_\mathrm{L}^2}B_\nu(T_{\mathrm{BB}}) + A_{\mathrm{PL}}\nu^{-0.75} 
\end{eqnarray}
where $d_\mathrm{L}$ is the luminosity distance and the three free parameters are the blackbody radius $R_{\mathrm{BB}}$, the blackbody temperature $T_{\mathrm{BB}}$, and the power law constant $A_{\mathrm{PL}}$. At the end of the fitting procedure, we obtain the best-fit $R_{\mathrm{BB}}$, $T_{\mathrm{BB}}$, and $A_{\mathrm{PL}}$.

During fitting, we followed \citet{Perley2019} and removed two clearly visible broad spectral features from the SEDs: the $Bbg$ data before $t = 11\,\mathrm{days}$ and the $Iz$ data after $t = 10\,\mathrm{days}$. The former is a broad absorption trough centered at approximately 4600\,\AA\,while, the latter is a broad emission bump around 9000\,\AA. We did not attempt to characterize these features in our models.

Figure \ref{fig:SEDfits} shows the best-fit models and observed SEDs at six sample epochs, including the three HST epochs. At early epochs, results from our fitting procedure are consistent with previous studies: (i) the UV-optical continuum was very consistent with a blackbody curve, (ii) the power law describes the excess IR emission reasonably well, and (iii) over time, the amplitude of the continuum decreased while the peak shifts from the NUV towards the optical. At the HST epochs, with precise UV photometry, we confirmed that the UV-optical continuum was still consistent with a blackbody curve (see $\chi^2/\mathrm{d.o.f}$ in Table \ref{tab:standardmodel}), i.e., optically thick. Furthermore, the spectral shape remained almost constant at these times, with a peak at $\lambda \sim 3000\,\mathrm{\AA}$. 

Based on the best-fit models, we also derived the UVOIR continuum bolometric luminosities of AT\,2018cow. The luminosity of the blackbody was taken simply as
\begin{equation}\label{eq:BBLum}
L_{\mathrm{BB}} = 4\pi R_{\mathrm{BB}}^2\sigma T_{\mathrm{BB}}^4
\end{equation}
where $\sigma$ is the Stefan-Boltzmann constant. The luminosity of the power law was calculated through 
\begin{equation}
L_{\mathrm{PL}} = 4\pi d_{\mathrm{L}}^2 A_{\mathrm{PL}}\left.\frac{\nu^{\alpha+1}}{\alpha+1}\right|_{\lambda=2\times10^{3}\,\mathrm{\AA}}^{\lambda=10^{7}\,\mathrm{\AA}}
\end{equation}
where $\alpha = -0.75$. Note that the integration range for the power law luminosity is arbitrarily chosen (covering most of the IR range) due to the lack of constraints, so the derived values should be treated with caution.

Finally, we note that it has also been suggested that dust could account for the excess IR emission \citep[e.g.,][]{Xiang2021,Metzger2022dust}. For this case, we checked to see if changing the power law to a dust model would impact the thermal properties derived from the blackbody. We performed fits in Appendix \ref{apdx:dustmodel} using a model in the form of \textsc{blackbody + dust} and found that the thermal properties derived from the blackbody are robust and largely unaffected by the description of the excess IR emission.

\begin{deluxetable*}{ccccccc}
\tablecaption{Properties of AT\,2018cow derived from the \textsc{blackbody + power law} model\label{tab:standardmodel}}
\tablewidth{0pt}
\tablehead{
\colhead{$t\,(\mathrm{day})$} & \colhead{$R_{\mathrm{BB}}\,(R_\odot)$} & \colhead{$T_{\mathrm{BB}}\,(\mathrm{K})$} & \colhead{$A_{\mathrm{PL}}\,(\mathrm{kJy}\,\mathrm{Hz}^{0.75})$} & \colhead{$L_{\mathrm{BB}}\,(10^{41}\,\mathrm{erg}\,\mathrm{s}^{-1})$} & \colhead{$L_{\mathrm{PL}}\,(10^{41}\,\mathrm{erg}\,\mathrm{s}^{-1})$} & \colhead{$\chi^2/\mathrm{d.o.f}$}
}
\startdata
3.09  & $10390.54^{+293.42}_{-292.89}$ & $32389.87^{+1200.87}_{-1093.39}$ & $42810.74^{+4923.90}_{-4787.35}$ & $4098.93^{+380.77}_{-328.24}$ & $40.44^{+4.65}_{-4.52}$ & 7.29\\
3.75  & $9791.83^{+261.03}_{-254.74}$  & $30239.58^{+1006.62}_{-945.66}$  & $46862.01^{+4272.03}_{-4149.33}$ & $2765.90^{+231.50}_{-206.24}$ & $44.26^{+4.04}_{-3.92}$ & 8.46\\
5.20  & $9069.67^{+218.83}_{-215.65}$  & $24888.00^{+723.03}_{-691.59}$   & $31603.12^{+1608.62}_{-1610.10}$ & $1087.97^{+76.51}_{-70.27}$   & $29.85^{+1.52}_{-1.52}$ & 3.24\\
6.18  & $7312.00^{+188.23}_{-188.46}$  & $26147.57^{+767.39}_{-742.33}$   & $25514.71^{+1836.36}_{-1908.38}$ & $861.17^{+60.96}_{-54.51}$    & $24.10^{+1.73}_{-1.80}$ & 6.98\\
6.57  & $7409.44^{+201.46}_{-191.95}$  & $24450.06^{+766.24}_{-752.87}$   & $21024.99^{+1702.87}_{-1616.00}$ & $676.87^{+49.88}_{-47.63}$    & $19.86^{+1.61}_{-1.53}$ & 6.93\\
8.26  & $7398.53^{+205.56}_{-194.68}$  & $20860.37^{+550.49}_{-549.81}$   & $27343.18^{+1823.33}_{-1851.81}$ & $357.51^{+20.06}_{-19.59}$    & $25.83^{+1.72}_{-1.75}$ & 12.72 \\
9.05  & $6780.06^{+164.11}_{-161.35}$  & $20656.62^{+495.64}_{-452.31}$   & $23496.85^{+1396.62}_{-1404.06}$ & $288.65^{+14.72}_{-12.66}$    & $22.19^{+1.32}_{-1.33}$ & 13.81\\
10.04 & $6057.73^{+154.69}_{-158.02}$  & $21059.10^{+526.03}_{-515.52}$   & $18246.62^{+1211.06}_{-1224.30}$ & $248.71^{+13.34}_{-12.41}$    & $17.24^{+1.14}_{-1.16}$ & 7.76 \\
13.56 & $5801.13^{+177.65}_{-166.56}$  & $18053.47^{+387.80}_{-379.63}$   & $12662.24^{+539.95}_{-523.22}$   & $123.27^{+4.53}_{-4.12}$      & $11.96^{+0.51}_{-0.49}$ & 1.71 \\
14.53 & $5471.49^{+188.58}_{-166.53}$  & $17694.89^{+392.56}_{-403.78}$   & $10691.38^{+462.76}_{-439.73}$   & $101.18^{+3.83}_{-3.41}$      & $10.10^{+0.44}_{-0.42}$ & 1.00 \\
15.54 & $5344.10^{+232.54}_{-219.13}$  & $16818.06^{+446.58}_{-423.34}$   & $10933.35^{+514.38}_{-529.27}$   & $78.80^{+2.95}_{-2.56}$       & $10.33^{+0.49}_{-0.50}$ & 1.57 \\
21.29 & $3542.78^{+108.92}_{-110.16}$  & $16072.67^{+396.12}_{-369.71}$   & $10824.82^{+544.97}_{-571.80}$   & $28.87^{+1.21}_{-1.05}$       & $10.22^{+0.51}_{-0.54}$ & 9.46 \\
22.28 & $3381.20^{+129.20}_{-126.69}$  & $15719.83^{+428.84}_{-401.72}$   & $9542.58^{+729.34}_{-712.72}$    & $24.08^{+0.91}_{-0.88}$       & $9.01^{+0.69}_{-0.67}$  & 12.92 \\
23.37 & $3014.36^{+154.47}_{-152.37}$  & $16331.95^{+574.30}_{-536.04}$   & $12688.59^{+964.25}_{-943.23}$   & $22.31^{+0.99}_{-0.96}$       & $11.99^{+0.91}_{-0.89}$ & 14.69 \\
28.22 & $3055.34^{+139.35}_{-140.42}$  & $13575.65^{+423.38}_{-379.05}$   & $9216.13^{+514.69}_{-548.52}$    & $10.95^{+0.43}_{-0.39}$       & $8.71^{+0.49}_{-0.52}$  & 8.09  \\
30.17 & $2338.37^{+139.42}_{-143.08}$  & $14631.13^{+654.20}_{-534.25}$   & $6019.21^{+492.78}_{-510.99}$    & $8.67^{+0.51}_{-0.43}$        & $5.69^{+0.47}_{-0.48}$  & 8.89 \\
34.19 & $1404.86^{+132.66}_{-123.36}$  & $17659.87^{+1328.23}_{-1159.81}$ & $4369.81^{+399.32}_{-373.34}$    & $6.62^{+0.85}_{-0.67}$        & $4.13^{+0.38}_{-0.35}$  & 2.34 \\
36.27 & $2010.08^{+138.44}_{-135.02}$  & $14522.74^{+808.41}_{-724.78}$   & $3478.06^{+110.78}_{-110.37}$    & $6.20^{+0.67}_{-0.58}$        & $3.29^{+0.10}_{-0.10}$  & 9.89 \\
43.84 & $998.05^{+115.05}_{-109.48}$   & $18023.01^{+1847.57}_{-1513.66}$ & $2533.21^{+243.13}_{-237.60}$    & $3.63^{+0.66}_{-0.51}$        & $2.39^{+0.23}_{-0.22}$  & 2.63 \\
50.25 & $1314.05^{+47.20}_{-53.26}$    & $14121.14^{+200.01}_{-190.78}$   & $299.12^{+227.85}_{-186.33}$     & $2.37^{+0.07}_{-0.09}$        & $0.28^{+0.22}_{-0.18}$  & 5.27 \\
55.87 & $828.62^{+56.45}_{-58.61}$     & $15868.78^{+465.19}_{-421.43}$   & $579.05^{+204.64}_{-190.43}$     & $1.50^{+0.07}_{-0.07}$        & $0.55^{+0.19}_{-0.18}$  & 7.15 \\
60.11 & $793.56^{+50.75}_{-51.30}$     & $15541.62^{+429.19}_{-388.80}$   & $319.48^{+140.67}_{-135.18}$     & $1.27^{+0.05}_{-0.05}$        & $0.30^{+0.13}_{-0.13}$  & 7.66
\enddata
\tablecomments{Errors given here are statistical errors from the 15.9th and 84.1th percentile. The degrees-of-freedom ($d.o.f$) is taken to be the number of data points minus the number of free model parameters.}
\end{deluxetable*}

\begin{figure}
\epsscale{1.0}
\plotone{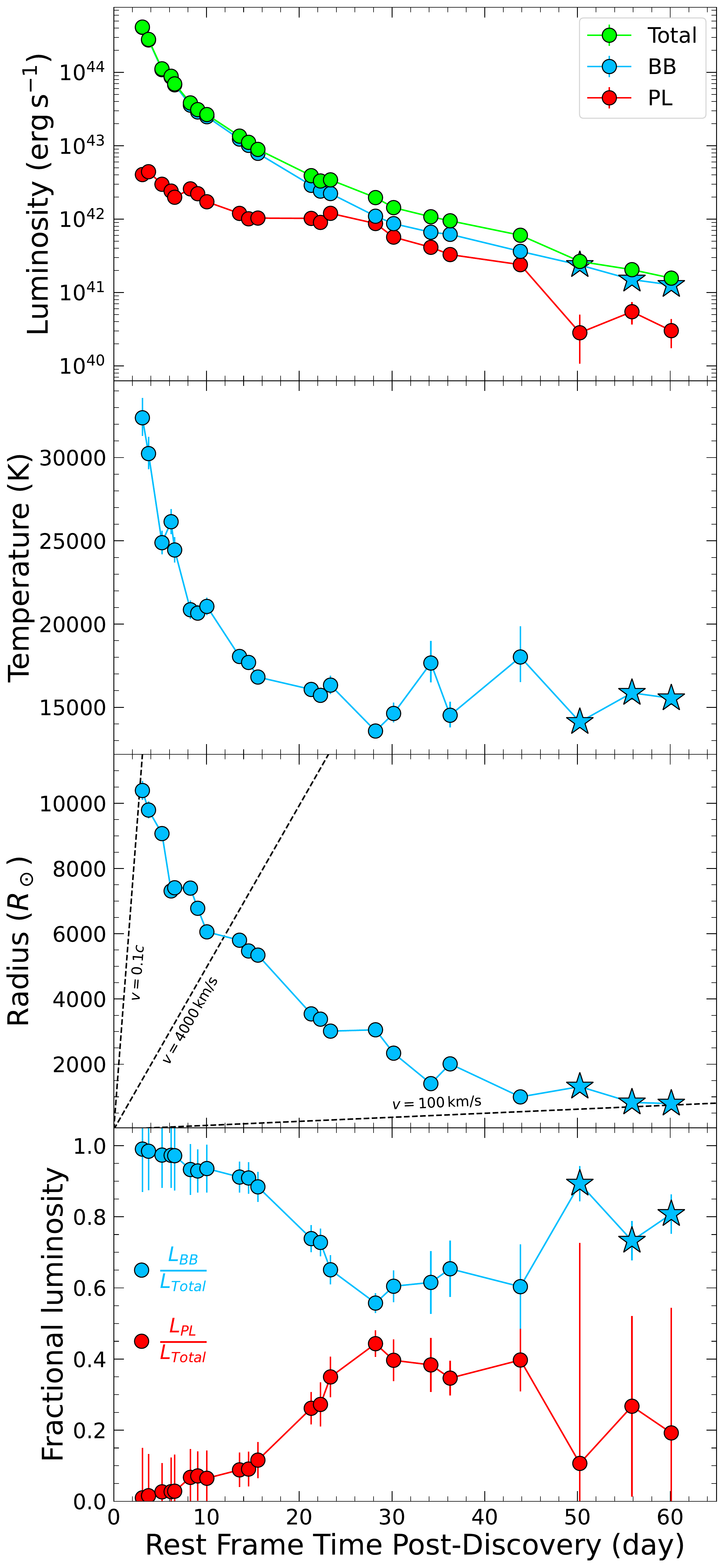}
\caption{Properties derived from fitting \textsc{blackbody + power law} model. Properties derived using HST photometry are shown as stars. The blackbody luminosity was calculated through the Stefan-Boltzmann law while the power law luminosity was calculated by integrating between $2000\,\mathrm{\AA}\leq\lambda\leq1\,\mathrm{mm}$.
\label{fig:standardresults}}
\end{figure}

\begin{figure*}
\epsscale{1.0}
\plotone{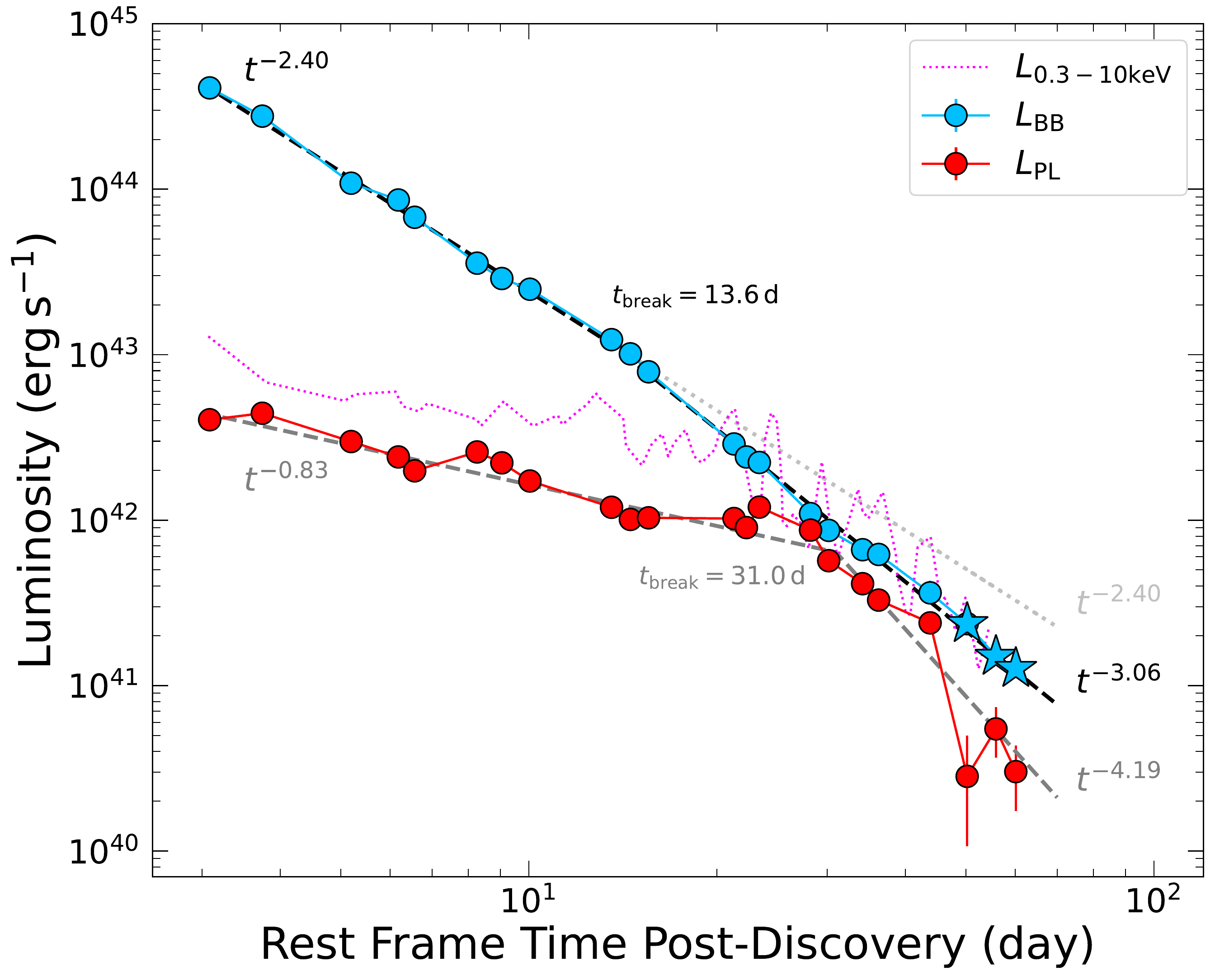}
\caption{The blackbody luminosity and power law luminosity ($L_{\mathrm{BB}}$ and $L_{\mathrm{PL}}$) derived from the \textsc{blackbody + power law} model, as well as the \emph{Swift} $0.3-10\,\mathrm{keV}$ luminosity from \citet{Margutti2019} are shown (similar X-ray light curves were also derived in \citealt{RiveraSandoval2018}, \citealt{Kuin2019}, and \citealt{Ho2019}). Blackbody luminosities derived using HST photometry are shown as stars. Best-fit broken power laws to the luminosities are shown in dashed lines. A dotted line is shown to illustrate the expected decline of the blackbody luminosity if the rate of decline stayed at $t^{-2.40}$.}
\label{fig:bolofit}
\end{figure*}

\subsection{Updated Thermal Properties ($t\sim3-60\,\mathrm{days}$)} \label{subsec:updatethermal}

The best-fit parameters, derived bolometric luminosities, and the reduced chi-square $\chi^2/d.o.f$ from fitting \textsc{blackbody + power law} at all 22 epochs are given in Table \ref{tab:standardmodel} and plotted over time in Figure \ref{fig:standardresults}. 

In general, our results at the earlier epochs are consistent with previous studies \citep[e.g.,][]{Perley2019}, namely the high peak luminosity ($L_{\mathrm{BB,peak}}\sim4\times10^{44}\,\mathrm{erg}\,\mathrm{s}^{-1}$) and peak temperature ($T_{\mathrm{BB,peak}}\sim30000\,\mathrm{K}$), the continuously receding blackbody radius, the temperature plateau at $T_{\mathrm{BB}}\sim15000\,\mathrm{K}$ after $t\sim15\,\mathrm{days}$, and the increasingly significant IR excess. At the later HST epochs, we found that the temperature remained high at $T_{\mathrm{BB}}\sim15000\,\mathrm{K}$ while the radius decreased to $R_{\mathrm{BB}}\lesssim1000\,R_\odot$, a size comparable to that of a red supergiant. These late thermal properties derived from the HST photometry are more precise compared to previous studies and suggest that the UV-optical continuum of AT\,2018cow at $t\simeq50-60\,\mathrm{days}$ comes from deeply-embedded hot thermalized material.

We further examined the bolometric light curves and performed simple fits to characterize their decline, shown in Figure \ref{fig:bolofit}. With the more precise late thermal properties, we report here a new finding: the luminosities decline faster at later epochs. Our initial attempt to model the decline with a single power law, following previous studies, yielded unsatisfactory fits ($\chi^2_{\mathrm{PL}}/d.o.f = 12.2$), and we instead found that a broken power law performed much better ($\chi^2_{\mathrm{bknPL}}/d.o.f = 2.8$). From fitting a broken power law, we found that the blackbody luminosity declined at a rate of $t^{-2.40}$ before $t_{\mathrm{break}}=13.6\,\mathrm{days}$, similar to the expectations of central engine scenarios where $L\propto t^{-\alpha}$ with $\alpha \approx 2-2.5$ \citep[][and references therein]{Margutti2019}. At $t>13.6\,\mathrm{days}$, the blackbody luminosity started declining at a faster rate of $t^{-3.06}$. The $t_{\mathrm{break}}=13.6\,\mathrm{days}$ is reminiscent of the transition point marked by the emergence of intermediate-width emission lines and the appearance of rapid X-ray variability. The decline in the blackbody luminosity at the later epochs is also very similar to the decline in the soft X-ray luminosity (Figure \ref{fig:bolofit}), which may indicate a correlation. We found a similar pattern for the power law luminosity, where the decline was initially gradual ($t^{-0.83}$) but became much faster ($t^{-4.23}$) after $t=31.5\,\mathrm{days}$. However, we note that the power law luminosity is poorly constrained due to the lack of IR data (see Figure \ref{fig:LC} and \ref{fig:SEDfits}) and ambiguity regarding the origin of the excess IR emission.

The updated thermal properties, derived from including the HST photometry, are summarized below.
\begin{itemize}
    \item At $t\simeq50-60\,\mathrm{days}$, the NUV-optical spectral shape is fully consistent with a blackbody that has a high temperature ($T_{\mathrm{BB}}\sim15000\,\mathrm{K}$) and a small radius ($R_{\mathrm{BB}}\lesssim1000\,R_\odot$). 
    \item The blackbody luminosity is better characterized by a broken power law that declined at a rate of $t^{-2.40}$ before a break at $t_{\mathrm{break}}=13.6\,\mathrm{days}$, and declined much faster at $t^{-3.06}$ after the break.
\end{itemize}
The combination of optically thick emission, high temperature, and rapid decline in blackbody luminosity at $t\simeq50-60\,\mathrm{days}$ is a unique feature of AT\,2018cow, even in the context of FBOTs, which we discuss in more detail in Section \ref{sec:promptinterpret}. 

\section{Constraints on the Power Source of the Fading Prompt Emission ($\lowercase{t}\sim20-60\,\mathrm{\lowercase{days}}$)} \label{sec:promptinterpret}

Through our analysis of the HST photometry, we showed that the UV-optical emission of AT\,2018cow at $t\simeq 50-60\,\mathrm{days}$ was very consistent with a blackbody that has a high temperature ($T_{\mathrm{BB}} \sim 15000\,\mathrm{K}$) and a small radius ($R_{\mathrm{BB}} \sim 1000\,R_\odot$). The natural interpretation of these results, given the lack of an observed nebular phase, is that the emission was still optically thick after two months and originated from thermalized material in the deeper regions. These thermal properties imply that there existed a power source (or multiple) that was still injecting energy at the later epochs. With the improved constraints, we also discovered that the blackbody luminosity declined faster after $t\sim15\,\mathrm{days}$, at a rate of $L_{\mathrm{BB}}\propto t^{-3.06}$, possibly associated with the evolution of the power source. 

In this section, we discuss the implications of our findings and place constraints on possible power sources of the fading prompt emission ($t\sim20-60\,\mathrm{days}$) of AT\,2018cow through comparisons with the literature and simple model estimates. We focus on three specific power sources generally favored by previous studies: radioactive decay (Section \ref{subsec:radioactivedecay}), ejecta-CSM interaction (Section \ref{subsec:promptCSM}), and central engine activity (Section \ref{subsec:promptengine}).

\subsection{Radioactive Decay} \label{subsec:radioactivedecay}

Generally, radioactive decay has been disregarded as the dominant power source of AT\,2018cow because an unphysical amount of $^{56}$Ni ($> 5\,M_\odot$) would be required to produce the peak luminosity \citep[][]{Margutti2019,Perley2019}. Still, the possible existence of a small amount of $^{56}$Ni has not been ruled out, which could have partially contributed to the observed thermal emission, especially at the later epochs. Constraining the mass of the $^{56}$Ni may also help distinguish possible progenitor systems and whether or not fallback is necessary to explain the lack of $^{56}$Ni.

Studies such as \citet{Xiang2021} and \citet{Pellegrino2022} have modeled previously published bolometric light curve of AT\,2018cow using a combination of ejecta-CSM interaction and radioactive decay and argued that radioactive decay could fully account for the emission after $t\sim25\,\mathrm{days}$. They inferred a $^{56}$Ni mass of $M_{\mathrm{Ni}}\sim0.1-0.3\,M_\odot$ from their models\footnote{Note that we do not quote the best-fit model from \citet{Pellegrino2022} but rather Model 20 because their best-fit model overestimates the emission after $t\sim50\,\mathrm{days}$.}.

\begin{figure}
\epsscale{1.0}
\plotone{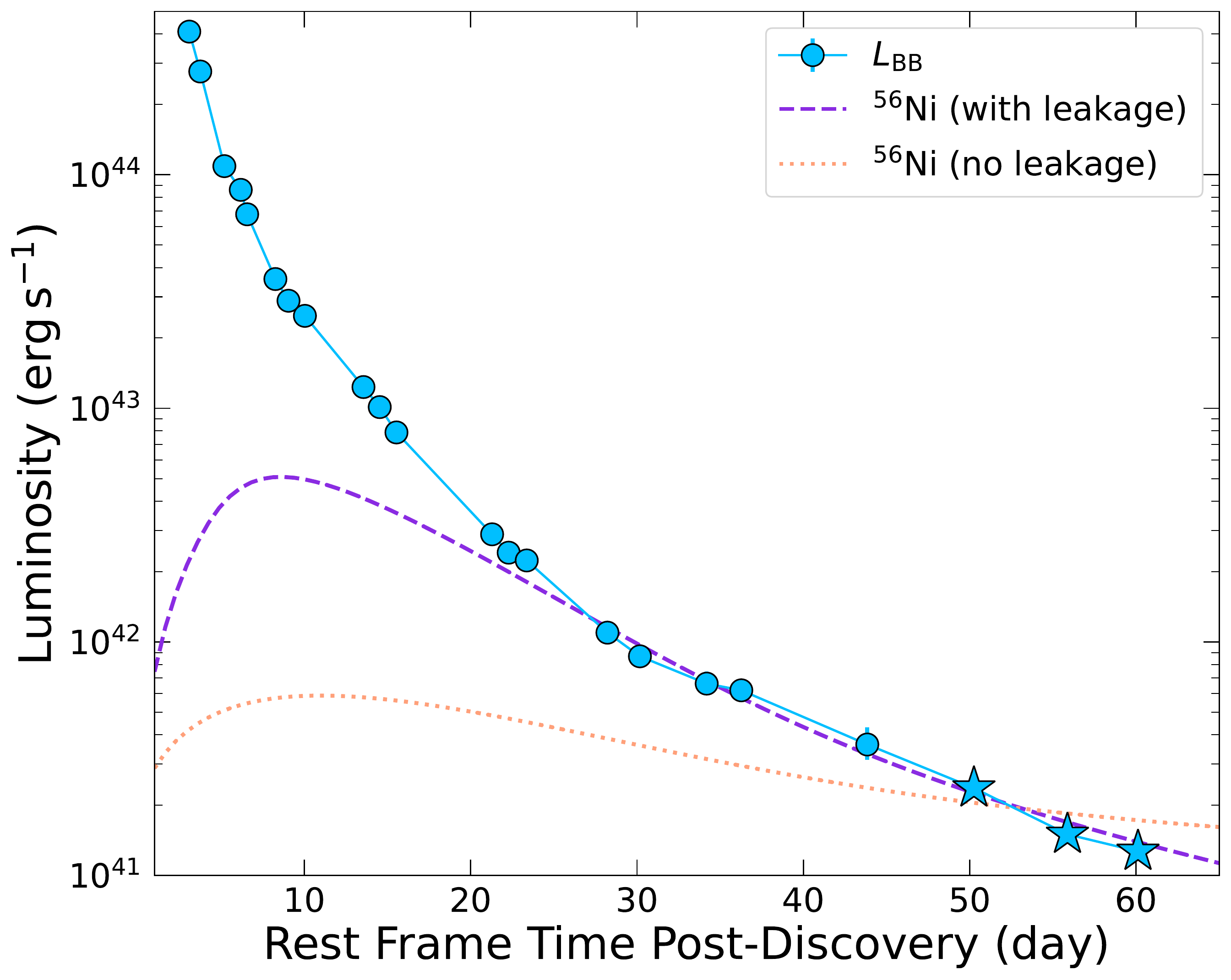}
\caption{The blackbody luminosity of AT\,2018cow ($L_{\mathrm{BB}}$) and the best-fit radioactive decay models ($^{56}$Ni) for the light curve after 25 days. The model without $\gamma$-ray leakage cannot fit the light curve at all while the model with $\gamma$-ray leakage can fully describe the light curve after $t\sim25\,\mathrm{days}$.
\label{fig:Nimodels}}
\end{figure}

We also modeled our updated bolometric light curve to investigate whether radioactive decay can power the late thermal emission and be consistent with the faster fading ($L_{\mathrm{BB}}\propto t^{-3.06}$) discovered in this study. Specifically, we used the semi-analytical model for radioactive decay from \citet{Chatzopoulos2012} that includes radiation diffusion to fit the thermal bolometric light curve at $t>25\,\mathrm{days}$. We considered two cases, one including $\gamma$-ray leakage through the factor $(1-e^{-(t/T_0)^{-2}})$, where $T_0$ is the trapping timescale, and the other assuming no leakage ($T_0 \rightarrow \infty$). For both cases, we adopted an optical opacity of $\kappa_{\mathrm{opt}}=0.2\,\mathrm{cm}^2\,\mathrm{g}^{-1}$.

Figure \ref{fig:Nimodels} shows the best-fit models and Table \ref{tab:Nimodels} gives the best-fit parameters. Without leakage, radioactive decay cannot reproduce the late-time light curve shape and even a small mass of $M_{\mathrm{Ni}}=0.019\,M_\odot$ overestimates the emission after $t\sim55\,\mathrm{days}$. For the no-leakage case, we also found an upper limit of $M_{\mathrm{Ni}}\lesssim0.013-0.019\,M_\odot$, which could power the exact luminosity at $t\simeq60\,\mathrm{days}$ assuming a range of $M_{\mathrm{ej}}=0.1-10\,M_\odot$ and $v_{\mathrm{ej}}=0.01c-0.1c$. On the other hand, when leakage is included, radioactive decay models can reproduce both the luminosity and observed decline at $t>25\,\mathrm{days}$ with a $M_{\mathrm{Ni}}=0.675^{+0.490}_{-0.335}\,M_\odot$ and a $T_0 = 9.54^{+3.90}_{-2.34}\,\mathrm{days}$.

Although the fit is reasonable when leakage is included, the inferred model parameters are inconsistent with other properties of AT\,2018cow. Note that our best-fit $^{56}$Ni mass is a few times larger than those inferred by \citet{Xiang2021} and \citet{Pellegrino2022}. This difference is likely due to the faster decline in our light curve at the later epochs: a faster decline can be produced by a larger $\gamma$-ray leakage (smaller trapping timescale), but a larger leakage reduces the heating and therefore requires a larger $^{56}$Ni mass. 
A $^{56}$Ni mass of $M_{\mathrm{Ni}}\sim0.3-1.1\,M_\odot$ is fairly difficult to explain given the lack of UV blanketing from iron peak elements in AT\,2018cow \citep[][]{Margutti2019}. This is especially true considering that both the small trapping timescale and high inferred velocity ($\sim$0.1c; higher than those observed in emission lines at similar epochs) required in this best-fit model could suggest that some fraction of the radioactive material is located in the outer portions of the ejecta. These inconsistencies imply that radioactive decay was unlikely the dominant power source over the entire $t\simeq25-60\,\mathrm{days}$ period.

Note that if radioactive decay is not required to start dominating at $t\simeq25\,\mathrm{days}$ but at a later date, then the required $M_{\mathrm{Ni}}$ can be smaller (e.g., $M_{\mathrm{Ni}}\lesssim0.013-0.019\,M_\odot$ would dominate at $t\gtrsim60\,\mathrm{days}$ assuming no leakage, as discussed above). 
Although small amounts of Nickel have been observed some SNe Ib/c with small ejecta masses, in most cases, line-blanketing is still observed. For example, in SN\,2011hs and SN\,2007Y ($\sim$0.02--0.03 M$_\odot$; \citealt{Prentice2016}) blanketing due to iron-peak elements suppressed the continuum at $\lambda < 5000$\,\AA\ to a level below that at longer wavelengths by $\sim$30--40 days post-discovery. Similar behavior was not detected for AT\,2018cow. Therefore, based on the constraints set by the lack of UV blanketing, we disfavor radioactive decay as a significant power source for the fading prompt emission of AT\,2018cow over $t\simeq25-60\,\mathrm{days}$, and we argue that alternative power sources are more likely.

\subsection{Ejecta-CSM Interaction}\label{subsec:promptCSM}

The bright radio synchrotron emission from AT\,2018cow has clearly revealed the existence of CSM around the progenitor \citep[][]{Margutti2019,Ho2019,Nayana2021}. Through radio observations, assuming a density distribution of $\rho = \dot{M}/(4\pi R^2v_\mathrm{w})$ with $v_\mathrm{w}=1000\,\mathrm{km}\,\mathrm{s}^{-1}$, \citet{Ho2019} inferred an $\dot{M}\approx4\times10^{-4}\,M_\odot\,\mathrm{yr}^{-1}$ within a radius of $R\lesssim1.7\times10^{16}\,\mathrm{cm}$ and \citet{Nayana2021} inferred an $\dot{M}\lesssim 4\times10^{-6}\,M_\odot\,\mathrm{yr}^{-1}$ at $R\gtrsim6\times10^{16}\,\mathrm{cm}$. The decline of $\dot{M}$ over distance suggests more enhanced mass loss closer to the explosion.

Several optical signatures, such as the featureless blue continuum, the narrow and intermediate-width lines, and the asymmetric line profiles, have also been used to argue the possibility that the thermal emission of AT\,2018cow was powered by ejecta-CSM interaction \citep[][]{Fox2019,Xiang2021,Pellegrino2022}. This scenario is quite plausible since a number of studies, through analytical and numerical treatments, have demonstrated that ejecta-CSM interaction can lead to bright rapidly-evolving SNe \citep[e.g.,][]{McDowell2018,Suzuki2019,Suzuki2020,Pellegrino2022,Maeda2022,Khatami2023}. Additionally, \citet{Fox2019} argued in favor of this scenario after finding similarities between AT\,2018cow and interacting SNe Ibn and IIn in their light curves and spectra. Following these arguments, studies such as \citet{Xiang2021} and \citet{Pellegrino2022} constructed models to show that ejecta-CSM interaction can power the optical peak of AT\,2018cow (and the subsequent emission up to $t\sim20\,\mathrm{days}$) through a dense shell of CSM with $\dot{M}\sim1\,M_\odot\,\mathrm{yr}^{-1}$ and an outer edge at $R_{\mathrm{out}}\sim1.0\times10^{14}\,\mathrm{cm}$. Note that the hypothetical CSM powering the optical peak is orders of magnitude denser than the CSM powering the radio synchrotron emission, which would suggest extreme eruption just before the explosion. Recently, \citet{Maund2023} reported flashes of optical polarization from AT\,2018cow at $t\sim5\,\mathrm{days}$ and $12\,\mathrm{days}$, hinting at an asymmetric configuration and possibly the existence of such dense CSM.

\begin{deluxetable}{ccc}
\tablecaption{Best-fit Radioactive Decay Models\label{tab:Nimodels}}
\tablewidth{0pt}
\tablehead{
\colhead{Parameters} & \colhead{With Leakage} & \colhead{No Leakage}}
\startdata
$M_{\mathrm{Ni}}$ ($M_\odot$) &  $0.675^{+0.490}_{-0.335}$  &  $0.0192^{+0.0005}_{-0.0005}$      \\
$v_{\mathrm{ej}}$ ($\mathrm{km}\,\mathrm{s}^{-1}$) & $5.37^{+7.12}_{-3.94}\times10^{4}$ & $1.71^{+0.91}_{-0.75}\times10^{3}$   \\
$M_{\mathrm{ej}}$ ($M_\odot$)  & $5.34^{+3.27}_{-3.63}$ &   $0.833^{+0.197}_{-0.262}$    \\
$R_0$ ($\mathrm{cm}$) &       $7.79^{+3.85}_{-4.04}\times10^{14}$        &  $11.04^{+2.14}_{-3.46}\times10^{14}$   \\
$t_0$ ($\mathrm{days}$) &       $-0.728^{+0.826}_{-0.793}$        &     $-1.89^{+0.17}_{-0.08}$          \\
$T_0$ ($\mathrm{days}$) &       $9.54^{+3.90}_{-2.34}$        &       $\infty$  
\enddata
\tablecomments{$M_{\mathrm{Ni}}$: $^{56}$Ni mass; $v_{\mathrm{ej}}$: ejecta velocity; $M_{\mathrm{ej}}$: ejecta mass; $R_0$: initial radius; $t_0$: initial time with respect to MJD 58285.441; $T_0$: $\gamma$-ray trapping timescale. Lower and upper errors are from the 15.9th and 84.1th percentile, respectively.}
\end{deluxetable}

Here, we do not construct another ejecta-CSM interaction model to describe our updated thermal properties. Instead, we discuss the implications of the observational properties through comparisons with existing studies of ejecta-CSM interaction. In particular, we discuss two points below: (1) it is unclear if current ejecta-CSM interaction models for AT\,2018cow can sufficiently explain the fading prompt emission at $t\gtrsim 20\,\mathrm{days}$, and (2) the peculiar thermal properties of AT\,2018cow are not naturally produced in general studies of aspherical ejecta-CSM interaction.

\subsubsection{Current Interaction Models for AT\,2018cow}

For AT\,2018cow, a great advantage of invoking CSM is that an aspherical configuration can qualitatively explain many of the peculiar observational properties. Often, studies will adopt a schematic that involves dense equatorial CSM (e.g., a disk) that leads to free-expanding fast ($v\sim0.1c$) ejecta near the polar region and slow-moving ($v\lesssim6000\,\mathrm{km}\,\mathrm{s}^{-1}$) ejecta-CSM interaction near the equator \citep[e.g., Figure 12 in][]{Margutti2019}. At an inclined line of sight, the initial peak of AT\,2018cow comes from the fast polar ejecta, and as this ejecta becomes optically thin, the photosphere recedes and reveals the embedded ejecta-CSM interaction along with intermediate-width emission lines at $t\sim15-20\,\mathrm{days}$. In this scenario, the embedded interaction powers the fading prompt emission over $t\sim20-60\,\mathrm{days}$ and maintains a temperature plateau at $T_{\mathrm{BB}}\sim15000\,\mathrm{K}$.

Although this schematic appears excellent qualitatively, a detailed ejecta-CSM interaction model that can reproduce the fading prompt emission of AT\,2018cow does not yet exist. In particular, it is unclear whether this scenario can really give rise to the combination of receding photosphere, optically thick emission, and rapid fading discovered in this study ($L\propto t^{-3}$ that persisted for more than a month after the emergence of the intermediate-width emission lines often associated with the embedded interaction). Many uncertainties remain, e.g., the discrepancy between $v\sim100\,\mathrm{km}\,\mathrm{s}^{-1}$ inferred from $R_{\mathrm{BB}}\sim1000\,R_\odot$ at $t\simeq60\,\mathrm{days}$ (Figure \ref{fig:standardresults}) and $v\sim3000-6000\,\mathrm{km}\,\mathrm{s}^{-1}$ inferred from the emission lines, and the observed rapid fading versus an expected slow fading due to high optical depth suppressing radiation. 

Current studies such as \citet{Xiang2021} and \citet{Pellegrino2022} have only modeled the optical peak of AT\,2018cow through spherical ejecta-CSM interaction and their choice of radioactive decay for the emission at $t\gtrsim25\,\mathrm{days}$ cannot explain the observed late thermal properties. Recently, \citet{Khatami2023} presented a general interaction framework involving a single spherical shell of CSM and was able to produce a rapid decline for AT\,2018cow in the edge breakout regime \citep[see Figure 14 in][]{Khatami2023}{}{} with a relatively low ejecta mass ($M_{\mathrm{ej}}\sim0.5\,M_\odot$) and supernova energy ($E_{\mathrm{sn}}\sim0.2\times10^{51}\,\mathrm{erg}\,\mathrm{s}^{-1}$). However, their current framework does not include spectral or photospheric evolution, and it is also not clear if the shock cooling phase under spherical symmetry can produce the peculiar thermal properties and optically thick emission of AT\,2018cow.

Lastly, we note that studies such as \citet{Metzger2022} and \citet{Lyutikov2022} have constructed more complex models for AT\,2018cow involving ejecta/wind-CSM interaction and a central engine. The interactions in these models are most likely aspherical, but similar uncertainties are present. In the model by \citet{Lyutikov2022}, the apparent receding photosphere was due to the material becoming optically thin over the first month, which is inconsistent with the observed optically thick emission at the end of the second month. The model by \citet{Metzger2022} does not explicitly track the receding photosphere. Therefore, it is unclear if the interaction in these specific models can result in the persistent receding photosphere and optically thick emission, and we instead turn to general simulations for more insights in Section \ref{subsubsec:asphericalCSM} below.

\subsubsection{Other Aspherical Ejecta-CSM Interactions} \label{subsubsec:asphericalCSM}

Since there is a lack of detailed aspherical ejecta-CSM interaction models in the context of AT\,2018cow, we turn to more general simulations involving aspherical CSMs for further insights into the observational properties. Note that these simulations typically examine transients with a timescale of $\sim$100s of days, and therefore our brief discussion here will be qualitative.

\citet{Suzuki2019} performed 2-D simulations involving disk CSMs with various masses and viewing angles and derived properties such as the color temperature and blackbody radius, allowing us to make some qualitative comparisons. In their simulations, the color temperature remained high for a long time, but most cases showed an increasing blackbody radius at all times. The only cases with an initial decline in blackbody radius were cases with an edge-on viewing angle, which had significantly prolonged light curves (slow rise and decline). The receding photosphere happens when the ejecta becomes transparent and reveals the disk, after which an expanding photosphere can be seen. These behaviors also demonstrate the effects of radiation diffusion: a large optical depth produces optically thick emission but suppresses the emission and prolongs the light curve. Therefore, general cases of disk interaction do not seem to easily produce the peculiar thermal properties of AT\,2018cow, namely the combination of optically thick emission, receding photosphere, and rapidly-declining brightness at the late epochs.

\citet{Kurfurst2020} performed 2-D simulations involving three CSM distributions: an equatorial disk, colliding wind shells, and a bipolar nebula. They derived two properties from their simulations that were also observed for AT\,2018cow: the emission line profiles and optical polarization degrees. Interestingly, they found that only interactions with colliding wind shells produced asymmetric emission lines. They also found an optical polarization degree of $\sim1-2\%$ for cases involving disk and bipolar CSM and $\sim0.01-0.5\%$ for cases involving colliding wind shells. In the case of AT\,2018cow, clear asymmetric lines and an optical polarization degree of $\sim0.3-2\%$ \citep[][]{SmithAtel2018,Maund2023} were observed besides the two high polarization flashes. This comparison suggests complex CSM distributions may need to be considered.

In conclusion, AT\,2018cow presents a unique opportunity to study ejecta-CSM interaction but there are still many open questions that are not answered with general simulations. Future detailed studies involving complex CSM distributions will be necessary to determine if ejecta-CSM interaction are capable of producing the combination of receding photosphere, optically thick emission, and rapidly declining luminosity. If ejecta-CSM interaction cannot offer satisfactory (quantitative) explanations for the fading prompt emission of AT\,2018cow at $t\sim20-60\,\mathrm{days}$, it could be an indication that a different power source (e.g., a central engine) dominated at this time.

\subsection{Central Engine -- Wind-Reprocessed Framework} \label{subsec:promptengine}

Central engines (NSs and BHs) are alternative power sources often invoked to explain FBOTs because they can release an enormous amount of energy over a short timescale \citep[e.g.,][]{Yu2015,2015MNRAS.451.2656K}. Central engines are particularly favored for the case of AT\,2018cow because they can also naturally explain the high ejecta energy, low ejecta mass, and persistent X-ray emission \citep[][]{Margutti2019,Ho2019}. Currently, a variety of central engine scenarios have been proposed to explain AT\,2018cow, typically involving jets \citep[][]{Soker2019,Gottlieb2022,Soker2022} or winds driven by the engine \citep[][]{Yu2019,Lyutikov2019,Piro2020,Uno2020,Lyutikov2022,Metzger2022} that may interact with pre-existing CSM. Specific scenarios involving TDEs with an IMBH \citep[with $M_{\mathrm{BH}} \lesssim 10^5\,M_\odot$; see review by][]{Greene2019IMBHReview} or a SMBH (with $M_{\mathrm{BH}} \gtrsim 10^5-10^6\,M_\odot$) have also been proposed \citep{Perley2019,Kuin2019}. The proposed scenarios also involve different progenitor systems, such as single or double WD systems producing NSs \citep[][]{Yu2019,Lyutikov2019,Lyutikov2022} or common envelope evolution involving the engine \citep[][]{Soker2019,Metzger2022,Soker2022}. 
Moving forth, to understand the nature of AT\,2018cow and similar transients, it is crucial to distinguish different scenarios.

\begin{figure}
\epsscale{1.15}
\plotone{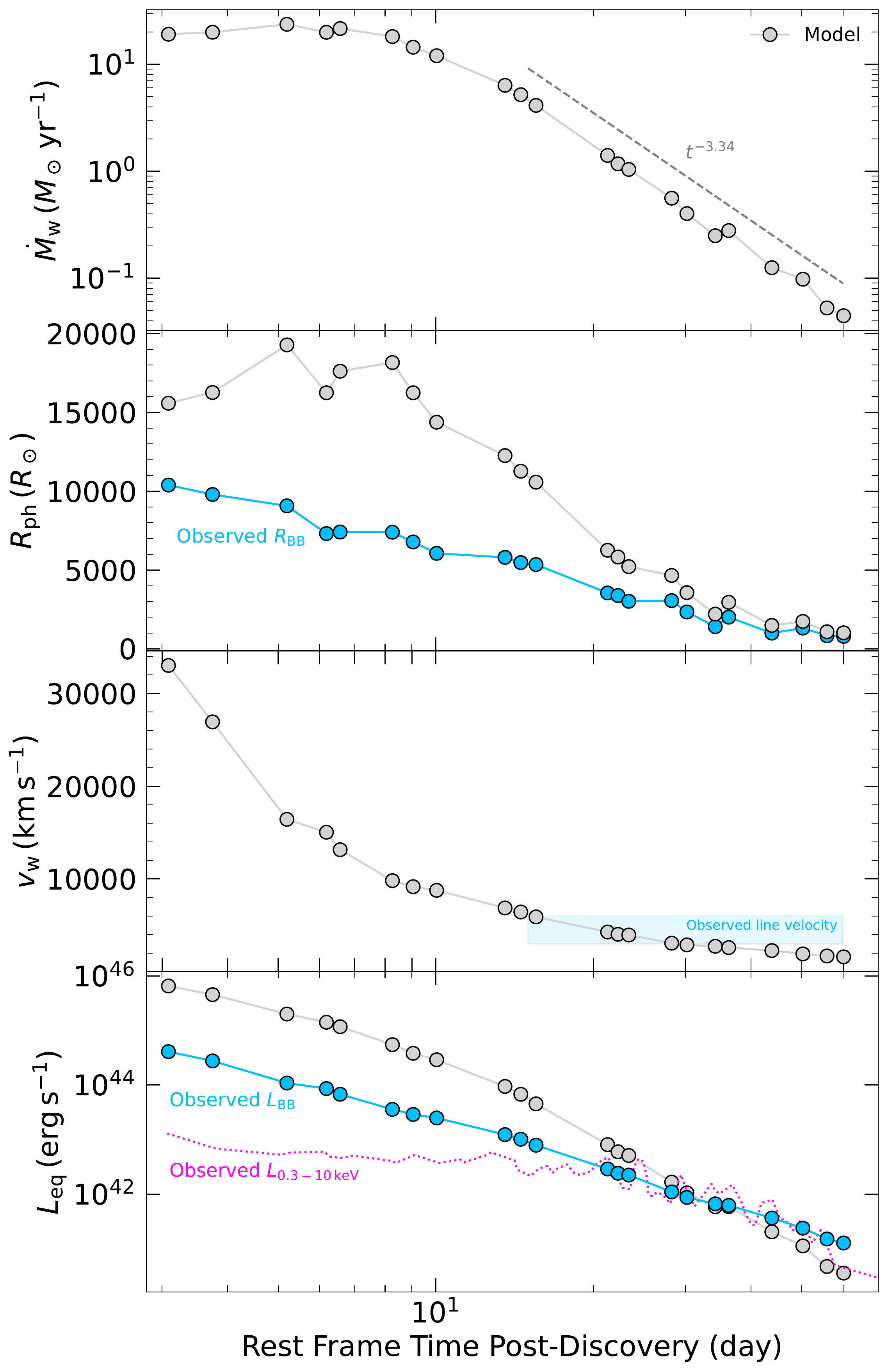}
\caption{The wind mass-loss rate $\dot{M}_{\mathrm{w}}$ (top panel), photosphere radius $R_{\mathrm{ph}}$ (second panel), wind velocity $v_{\mathrm{w}}$ (third panel), and luminosity at equipartition radius $L_{\mathrm{eq}}$ (bottom panel) derived using observed properties of AT\,2018cow based on the wind-reprocessed transient model from \citet{Uno2020}. A power law fit to the wind mass-loss rate after $t\sim15\,\mathrm{days}$ is also shown on the top panel. A shaded region is shown on the third panel to indicate the observed emission line velocity $3000\,\mathrm{km}\,\mathrm{s}^{-1}\lesssim v \lesssim 6000\,\mathrm{km}\,\mathrm{s}^{-1}$. Observed properties are also plotted for comparisons with $R_{\mathrm{ph}}$ and $L_{\mathrm{eq}}$.
\label{fig:UnoModel}}
\end{figure}

Often, models will use the rise time and peak optical luminosity to constrain the configurations of the engines. For AT\,2018cow, the late thermal properties can also provide additional constraints if the fading prompt emission was associated with a central engine. In particular, the receding photosphere and optically thick emission have significant implications on the evolution of the optical depth and cannot be explained by a simple spherically-expanding engine-powered ejecta. The scenario mostly likely require reprocessing involving some form of wind driven by the engine. Specific configurations of the wind may be able to explain the large optical depth, receding photosphere, and appearance of asymmetric intermediate-width lines. The observed rate of decline in blackbody luminosity at the later epochs ($t^{-3}$) may also constrain the input power of the engine.

Here, we consider an example case and show that a scenario involving wind driven by a central engine can reasonably explain the late thermal properties of AT\,2018cow presented in this study. Specifically, we characterize AT\,2018cow under the wind-reprocessed framework constructed by \citet{Piro2020} and \citet{Uno2020} \citep[see also][for numerical simulation]{Calderon2021}. This framework involves continuous outflow/wind driven by a central engine and, following the formulation of \citet{Uno2020}, relates the observed luminosity and temperature with wind mass-loss rate $\dot{M}_{\mathrm{w}}$, wind (launch) velocity $v_{\mathrm{w}}$, and photosphere radius $R_{\mathrm{ph}}$. Both \citet{Piro2020} and \citet{Uno2020} have modeled AT\,2018cow but using previously derived thermal properties \citep[e.g., from][]{Perley2019}. We repeat this modeling to see if our updated properties introduce different interpretations. 

Figure \ref{fig:UnoModel} shows the model properties derived for AT\,2018cow using the formulation from \citet{Uno2020}. Note that we ignore possible interactions with pre-existing CSM because of the uncertainties discussed in Section \ref{subsec:promptCSM}, and also because we are interested in seeing if wind can account for the late-time observational properties without interaction with CSM. Also, since our focus is on the late properties, we will not comment on the performance of the model before $t\sim15\,\mathrm{days}$. This early emission may or may not be powered by the same mechanism.

The model wind velocity declines steadily after $t\sim15\,\mathrm{days}$ and is broadly consistent with the velocity derived from the intermediate-width lines that appeared after $t\sim15\,\mathrm{days}$ ($3000\,\mathrm{km}\,\mathrm{s}^{-1}\lesssim v \lesssim 6000\,\mathrm{km}\,\mathrm{s}^{-1}$). The model photosphere radius also shows a steady decline after $t\sim15\,\mathrm{days}$, consistent with the observed receding photosphere. We note that although the photosphere and blackbody radii are similar at these epochs, the two are not necessarily comparable because the model photosphere radius takes electron scattering into account. The wind mass-loss rate declines as a power law after $t\sim15\,\mathrm{days}$ at a rate of $\dot{M}_{\mathrm{w}} \propto t^{-3.35}$. Note that this decline rate is much faster compared to the approximate rate found by \citet{Uno2020} and \citet{Piro2020}, $\dot{M}_{\mathrm{w}} \propto t^{-5/3}$, which was consistent with the expected accretion rate for a fallback scenario. The faster decline rate in $\dot{M}_{\mathrm{w}}$ is likely due to the faster decline rate in the blackbody luminosity found in this study. Lastly, we also calculated the luminosity $L_{\mathrm{eq}}$ at the innermost equipartition radius (below which the internal and kinetic energy are in equipartition), which can be interpreted as the engine power. This luminosity is therefore given by $L_{\mathrm{eq}}=\dot{M}_{\mathrm{w}}v_{\mathrm{w}}^2/2$. As suggested by \citet{Piro2020}, the similarity between the engine power and the observed X-ray emission at the later epochs is consistent with the reprocessing picture.

Overall, the wind-reprocessed framework can reasonably produce the peculiar thermal properties associated with the fading prompt emission of AT\,2018cow. Even the asymmetric line profiles, as suggested by \citet{Uno2020}, can be explained by an asymmetric distribution of wind outflow. These results could suggest that wind driven by a central engine is sufficient, and ejecta-CSM interaction may not be necessary to account for the fading emission. Lastly, we note that our faster rate of $\dot{M}_{\mathrm{w}} \propto t^{-3.35}$ could have implications on the power output of the engine and the structure of the wind. For example, spherical wind outflow is assumed in these models, which may not be realistic. The actual scenario may have involved wind that was aspherical and outflow that decreased in size over time (i.e., smaller solid angle and lower brightness), which the model would have accounted for by decreasing the wind mass-loss rate.

\section{Summary \& Conclusion}\label{sec:conclusion}

In this study, we analyzed the first three HST observations of AT\,2018cow ($t \simeq 50.3,\,55.9,\,60.1\,\mathrm{days}$) to constrain the late thermal properties of the fading prompt emission. With significantly improved precision in the UV photometry, we confirmed that the fading prompt emission at $t\simeq50-60\,\mathrm{days}$ is blackbody (i.e., optically thick) with a high temperature ($T_{\mathrm{BB}}\sim15000\,\mathrm{K}$) and a small radius ($R_{\mathrm{BB}}\lesssim 1000\,R_\odot$). Furthermore, we found that although the blackbody luminosity initially declined at a rate of $t^{-2.40}$ \citep[similar with previous findings, e.g.,][]{Margutti2019,Perley2019}, the decline became much faster after $t_{\mathrm{break}}=13.6\,\mathrm{days}$, at a rate of $t^{-3.06}$. The combination of receding photosphere, high temperature, and optically thick emission with a rapidly declining luminosity is very peculiar and places significant constraints on possible power sources for the fading prompt emission. 

We disfavor radioactive decay as the dominant power source over $t\simeq25-60\,\mathrm{days}$ because significant $\gamma$-ray leakage is likely needed to produce the faster decline in luminosity, which drives up the required $^{56}$Ni mass to $M_{\mathrm{Ni}}\gtrsim0.3\,M_\odot$, inconsistent with the lack of UV line blanketing. Even assuming a $M_{\mathrm{Ni}}\sim0.01-0.02\,M_\odot$, which would be sufficient to start dominating at around $t\sim50-60\,\mathrm{days}$, the complete lack of blanketing in the HST SEDs is still quite difficult to explain given that the material is already partially optically thin as indicated by the receding photosphere. Therefore, the thermal emission observed at these later epochs is likely powered by an alternative power source.

Although we do not rule out ejecta-CSM interaction as a major contributor to the fading prompt emission, we argue that current models by \citet[][]{Xiang2021} and \citet{Pellegrino2022} are not sufficient in explaining the late thermal properties because of the assumption of spherical symmetry and the dependence on radioactive decay at $t\gtrsim25\,\mathrm{days}$. Similarly, although the shock cooling phase in the interaction model by \citet{Khatami2023} can match the rapid fading of AT\,2018cow, it is also not clear if this scenario can explain the spectral and photospheric evolution. Qualitative comparisons with general simulations involving aspherical CSMs \citep[][]{Suzuki2019,Kurfurst2020} show that the peculiar thermal properties of AT\,2018cow are not easily produced, namely the combination of receding photosphere, optically thick emission, and rapidly declining luminosity. The effects of optical depth in an expanding shock seems to create a major dilemma: material becoming fully optically thin can produce a receding photosphere and rapidly declining luminosity but cannot explain the optically thick emission, while optically thick material would significantly suppress the emission and prolong the light curve. Therefore, whether or not ejecta-CSM interaction can fully explain the late thermal emission of AT\,2018cow, and the exact CSM configuration required to do so remain open questions. 

On the other hand, we found that the wind-reprocessed framework \citep[][]{Piro2020,Uno2020} involving continuous outflow/wind driven by a central engine can reasonably explain the late thermal properties of AT\,2018cow. In this scenario, the photosphere is receding as earlier winds become optically thin, while the decline in luminosity can be associated with the decline in wind mass-loss rate $\dot{M}_\mathrm{w}$ and wind velocity $v_\mathrm{w}$ (following the model in \citealt{Uno2020}). Further supporting this scenario, we found the model $v_\mathrm{w}$ at $t\gtrsim 20\,\mathrm{days}$ to broadly match the inferred velocity from the intermediate-width lines $v\sim3000-6000\,\mathrm{km}\,\mathrm{s}^{-1}$. Finally, we found $\dot{M}_\mathrm{w}\propto t^{-3.34}$, which is significantly faster than the predicted fallback rate of $\dot{M}\propto t^{-5/3}$, perhaps implying deviation from standard fallback or asymmetric outflow. Overall, these findings suggest that a central engine, without the need to invoke ejecta-CSM interaction, may be sufficient in explaining the fading thermal emission of AT\,2018cow.

While the exact progenitor, explosion mechanism, and power source(s) of AT\,2018cow are still open to debate, constraints from the late-time HST measurements seem to be pointing towards an accreting central engine. Although an accreting central engine and the wind-reprocessed framework is not needed to explain most extragalactic transients, in the case of AT\,2018cow, the unique combination of bright X-ray emission, mildly-relativistic outflow, and persistent optically thick thermal emission can be more consistently explained by the central engine scenario. The continuously receding photosphere (typically not observed in most transients) further associates the observed late-time emission with central activities and itself is much easier to explain with continuous wind/outflow (implying continuous energy injection) rather than a single expanding shock. If this association is true, AT\,2018cow and other luminous FBOTs may form an entirely new class of transients powered predominantly by accretion that generates mildly relativistic outflow which powers the initial peak and continuous winds which power the optically thick fading emission. These transients, given the extreme accretion required, would likely involve BHs -- either newborn BHs from stellar collapse or (hypothetical) existing IMBH -- and bridge the gap between ultra-relativistic transients (e.g., Gamma-Ray Bursts) and non-relativistic transients (e.g., normal SNe). Future observations and additional theoretical works are necessary to confirm or rule out this hypothesis.

Our work also demonstrates the importance of late-time observations in providing additional constraints for unsettled cases such as AT\,2018cow. In the fortunate event that another ``Cow-like transient'' is discovered nearby similar to AT\,2018cow, late-time monitoring by HST and JWST can be highly beneficial in differentiating theoretical models and resolving the mysteries of these peculiar transients.

\begin{acknowledgments}

We thank the anonymous referee for comments and suggestions that improved this manuscript. We thank Natalie Ulloa, Nahir Mu$\tilde{\mathrm{n}}$oz-Elgueta, Abdo Campillay, Nidia Morrell, Jaime Vargas-Gonz\'{a}lez, and Jorge Anais Vilchez for carrying out the Swope observations. We thank Georgios Dimitriadis, David Jones, Wynn Jacobson-Gal\'{a}n, Karelle Siellez, and Yao Yin for helping with the spectroscopic observations and Paul Butler and Adriana Kuehnel for reducing the APF spectrum. Y.C. thanks Christopher D. Matzner for helpful discussions and Samantha Berek, Bolin Fan, Steffani Grondin, Ayush Pandhi, and Dang Pham for comments throughout this project.

Y.C. acknowledges support from the Natural
Sciences and Engineering Research Council of Canada (NSERC) Canada Graduate Scholarships -- Doctoral Program. M.R.D. acknowledges support from NSERC through grant RGPIN-2019-06186, the Canada Research Chairs Program, the Canadian Institute for Advanced Research (CIFAR), and the Dunlap Institute at the University of Toronto. C.D.K. is supported in part by a Center for Interdisciplinary
Exploration and Research in Astrophysics (CIERA) postdoctoral fellowship.

Parts of this research is based on observations made with the NASA/ESA Hubble Space Telescope obtained from the Space Telescope Science Institute, which is operated by the Association of Universities for Research in Astronomy, Inc., under NASA contract NAS 5–26555. The HST observations are associated with program 15600 and can be accessed via \dataset[10.17909/5ry6-7r85]{http://dx.doi.org/10.17909/5ry6-7r85}. We acknowledge the use of public data from the Swift data archive. 

A subset of the data presented herein were obtained at the W.\ M.\ Keck Observatory, which is operated as a scientific partnership among the California Institute of Technology, the University of California, and the NASA. The Observatory was made possible by the generous financial support of the W.\ M.\ Keck Foundation. The authors wish to recognise and acknowledge the very significant cultural role and reverence that the summit of Maunakea has always had within the indigenous Hawaiian community. We are most fortunate to have the opportunity to conduct observations from this mountain. Keck observations were conducted on the stolen land of the k\={a}naka `\={o}iwi people. We stand in solidarity with the Pu'uhonua o Pu'uhuluhulu Maunakea in their effort to preserve these sacred spaces for native Hawai`ians. 

Shane 3~m and Automated Planet Finder 2.4~m telescope observations at the Lick Observatory were conducted on the stolen land of the Ohlone (Costanoans), Tamyen, and Muwekma Ohlone tribes. A major upgrade of the Kast spectrograph on the Shane 3~m telescope at Lick Observatory was made possible through generous gifts from the Heising-Simons Foundation as well as William and Marina Kast. We especially would like to thank Ken and Gloria Levy for their generous contribution that helped fund completion of the Levy spectrometer. Research at Lick Observatory is partially supported by a generous gift from Google.

\end{acknowledgments}

\vspace{5mm}
\facilities{HST (UVIS), Swift (UVOT), Swope, Shane (Kast), APF (Levy), Keck:I (LRIS).}

\software{Astropy \citep{2013A&A...558A..33A,2018AJ....156..123A}, \texttt{emcee} \citep{2013PASP..125..306F}, {\tt dolphot} \citep{dolphot}, \texttt{hotpants} \citep{becker2015}, HEASoft, SciPy \citep{2020scipy}.}

\appendix

\section{Additional Optical Spectroscopy}\label{apdx:spectra}

Here we present a set of previously unpublished optical spectra of AT\,2018cow, obtained with the Kast dual-beam spectrograph \citep{KAST} on the Lick Shane 3~m telescope, the Levy \citep{APF} spectrograph on the 2.4~m Automated Planet Finder (APF) telescope and the Low-Resolution Imaging Spectrograph \citep[LRIS;][]{LRIS} on the 10~m Keck~I telescope. In each epoch, we aligned the slit to the parallactic angle to minimize the effects of atmospheric dispersion \citep{Filippenko1982}. A log of all spectroscopic observations is presented in Table~\ref{tab:speclog}.

The Kast and LRIS data were reduced using a custom data reduction pipeline\footnote{\url{https://github.com/msiebert1/UCSC\_spectral\_pipeline}} \citep{Siebert2019}. The two-dimensional spectra were bias-corrected, flat-field corrected, adjusted for varying gains across different chips and amplifiers, and trimmed. Cosmic-ray rejection was applied using the {\tt pzapspec} algorithm to individual frames. Multiple frames were then combined with appropriate masking. One-dimensional spectra were extracted using the optimal algorithm \citep{Horne1986}. The spectra were wavelength-calibrated using internal comparison-lamp spectra with linear shifts applied by cross-correlating the observed night-sky lines in each spectrum to a master night-sky spectrum. Flux calibration was performed using spectro-photometric standard stars at a similar airmass to that of the science exposures, with ``blue'' (hot subdwarfs; i.e., sdO) and ``red'' (low-metallicity G/F) standard stars and corrected for atmospheric extinction. By fitting the continuum of the flux-calibrated standard stars, we determined the telluric absorption in those stars and applied a correction, adopting the relative airmass between the standard star and the science image to determine the relative strength of the absorption. We allowed for slight shifts in the telluric A and B bands, which we determined through cross-correlation. Finally, we combined the calibrated one-dimensional spectra using a $\sim$100\AA\ overlap region between the red and blue sides. 

The APF data was reduced using a custom raw reduction package developed for Iodine based precision velocity spectrometers by the APF team. The reduction package performs standard flat fielding, scattered light subtraction, order tracing, and cosmic ray removal. The spectra are wavelength calibrated against the NIST Fourier Transform Spectrometer spectral atlas of the APF Iodine cell, with a resolution of 1,000,000 and a S/N $\sim$1,000. Telluric lines are identified by comparing the spectrum of a rapidly rotating B star to a synthetic telluric atlas. With the exception of telluric lines, the spectrum of the rapidly rotating B star is nearly featureless. Pixels with telluric lines are given zero weight for the wavelength and velocity determination. The final spectrum is presented in vacuum wavelengths.

\begin{figure}
\includegraphics[width=0.47\textwidth]{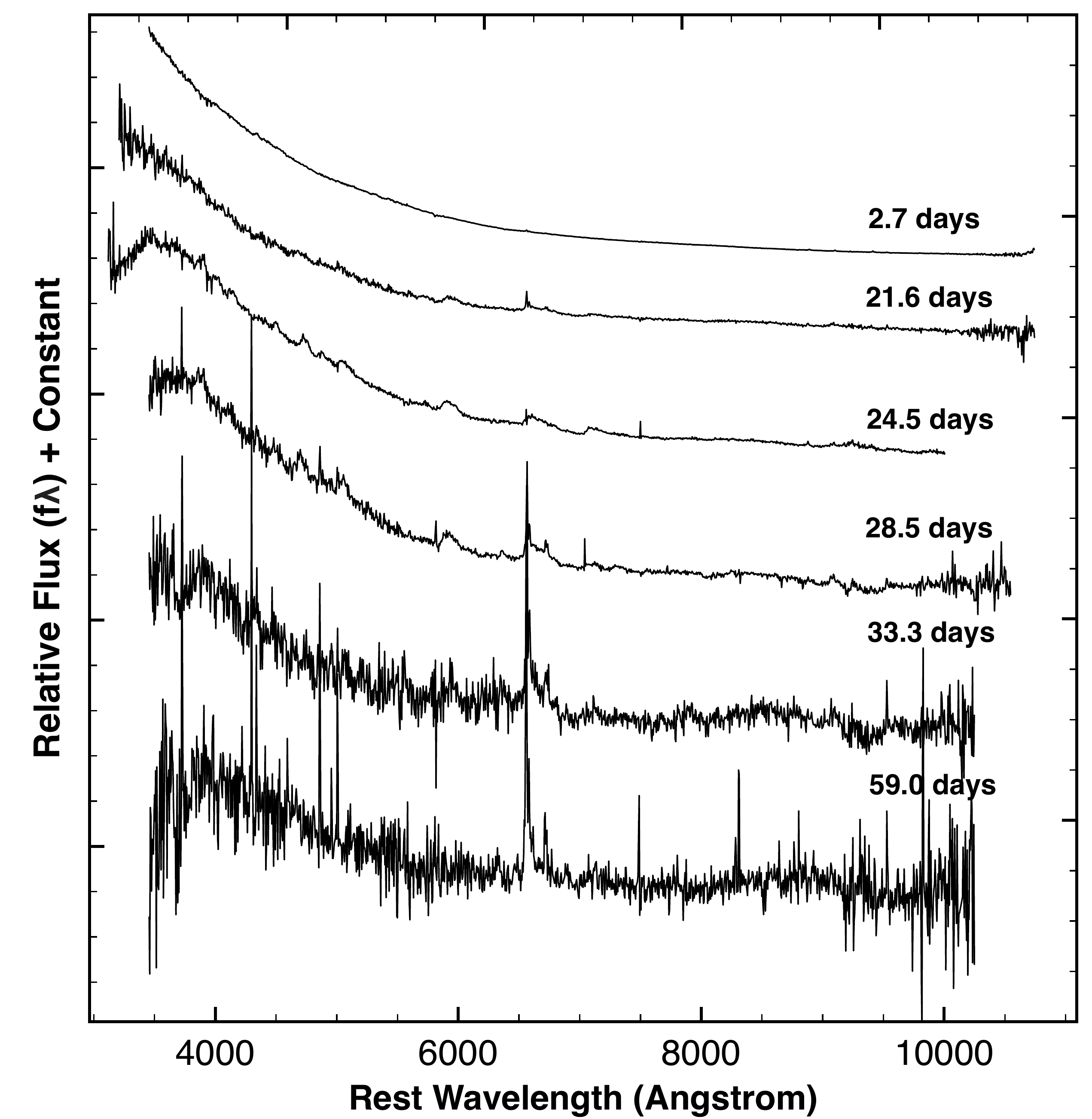}
\caption{Kast and LRIS spectra of AT\,2018cow over $t\simeq2-59\,\mathrm{days}$, showing an initial featureless spectrum with emerging intermediate-width lines at $t\gtrsim 15\,\mathrm{days}$.}\label{fig:spec-lowres}
\end{figure}

A summary plot of all low-resolution spectra are given in Figure~\ref{fig:spec-lowres}. Generally, our spectra are consistent with previously published ones \citep[e.g.,][]{Perley2019,Margutti2019,Xiang2021}, showing a featureless continuum at early epochs and emerging intermediate-width lines of hydrogen, helium, and other species after $t\sim15\,\mathrm{days}$. From the lower resolution spectra taken at $t\simeq21-33\,\mathrm{days}$, we found the intermediate-width lines to be redshifted by $\sim$3000\,km\,s$^{-1}$ and measured the velocities to be $v\sim5000-7000\,\mathrm{km}\,\mathrm{s}^{-1}$, consistent with previously reported values. Unfortunately, the spectrum taken at $t=58.98\,\mathrm{days}$ (contemporaneous with the late-time HST observations) did not have sufficient S/N for extracting reliable line velocities from the intermediate-width lines.

\begin{table*}
\centering
\caption{Log of spectroscopic observations for AT~2018cow.}
\label{tab:speclog}
\begin{tabular}{ccccccc}
\hline
\hline
MJD & t (days) & Telescope & Instrument & R & Setup & Exp. Time (s)\\
\hline
58288.190 & 2.71 & Shane  & Kast  & 400      & 452/3306, 300/7500 & 2$\times$450, 2$\times$435 \\ 
58290.283 & 4.77 & APF    & Levy  & 110,000 & 41 gr/mm R-4 Echelle & 6$\times$1200 \\
58307.313 & 21.57 & Shane  & Kast  & 400      & 452/3306, 300/7500 & 1845, 3$\times$600 \\ 
58310.324 & 24.54 & Keck I   & LRIS & 400       & 600/4000, 400/8500 & 1200, 2$\times$554 \\ 
58314.301 & 28.46 & Shane  & Kast  & 400       & 452/3306, 300/7500 & 4$\times$1845, 6$\times$600 \\ 
58314.347 & 28.50 & Shane  & Kast  & 10,200  & 1200/5000 & 6$\times$600\\ 
58319.266 & 33.35 & Shane  & Kast  & 400      & 452/3306, 300/7500 & 2$\times$1845+1$\times$1925, 9$\times$600 \\ 
58345.248 & 58.98 & Shane  & Kast  & 400      & 452/3306, 300/7500 & 3$\times$1845, 9$\times$600\\ 
\hline
\end{tabular}
\end{table*}

\begin{figure}
\includegraphics[width=0.47\textwidth]{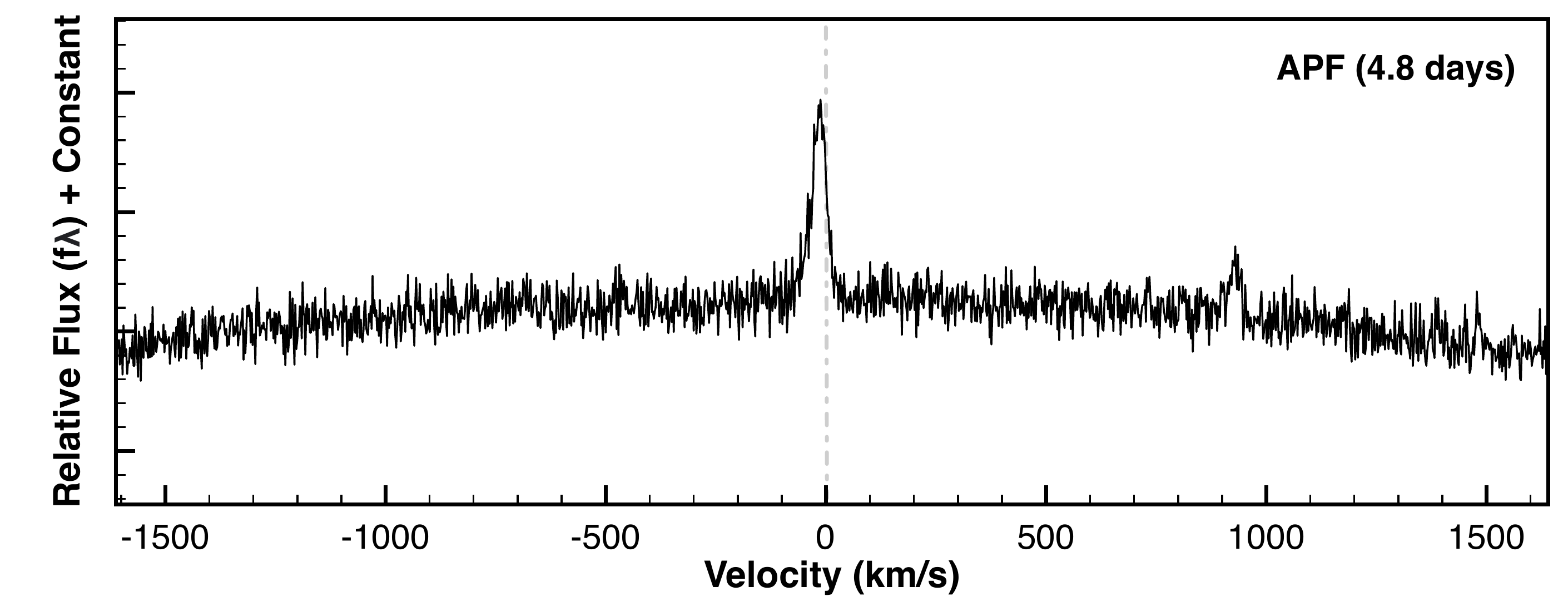}
\includegraphics[width=0.47\textwidth]{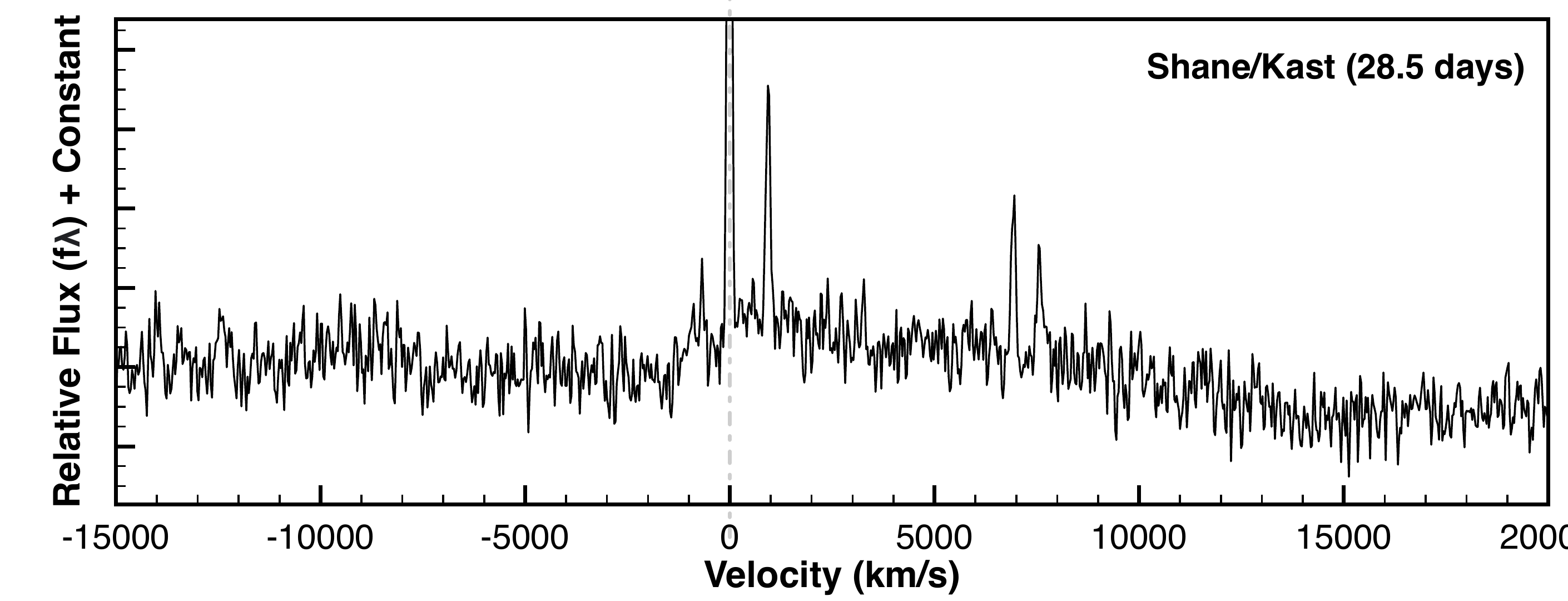}
\caption{High-resolution APF and Kast spectra in velocity space centered at H$\alpha$ in rest frame relative to host redshift.}\label{fig:spec-highres}
\end{figure}

In the high-resolution APF spectrum taken at $t=4.77\,\mathrm{days}$, only narrow H$\alpha$ and [N\,\textsc{II}] $\lambda$6583 emission lines were resolved with a width of $v\sim 30-40\,\mathrm{km}\,\mathrm{s}^{-1}$ (see Figure~\ref{fig:spec-highres}). These narrow features are also blueshifted by $\sim$15\,km\,s$^{-1}$ relative to the rest wavelength, consistent with the gas velocity at the site of AT\,2018cow \citep[see Figure 6 in][]{Lyman2020}. Therefore, the Levy spectrum confirms that no narrow lines from the transient existed at $t=4.77\,\mathrm{days}$ (and possibly for the first few days), as the detected lines are likely associated with the host galaxy.

\section{Dust Model for the Excess IR Emission}\label{apdx:dustmodel}

We introduced a model of the form \textsc{blackbody + dust} to test the robustness of the blackbody properties, i.e., whether or not they change if the power law is switched to a dust model.

We adopted a warm dust model assuming an ideal cloud of spherical dust grains with uniform size, composition, and temperature, with a flux density given by \citep[Equation (4) of ][]{1983QJRAS..24..267H}:
\begin{equation}
    F_{\nu,\mathrm{d}}(M_\mathrm{d}, T_\mathrm{d}) = \frac{3Q_\nu(a)}{4\rho a}\frac{M_\mathrm{d}}{d_\mathrm{L}^2}B_\nu(T_\mathrm{d}) 
\end{equation}
where $Q_\nu(a)$ is the dust absorption coefficient, $\rho$ is the mass density of the grain, $a$ is the grain radius, $M_\mathrm{d}$ is the total mass of the dust cloud, and $T_\mathrm{d}$ is the temperature of the dust cloud. We assumed a fiducial dust size of of $a=0.1\,\mu\mathrm{m}$. We used the coefficients $Q_\nu(a)$ derived in \citet{1984ApJ...285...89D} and \citet{1993ApJ...402..441L} from Draine’s website\footnote{\url{https://www.astro.princeton.edu/~draine/dust/dust.diel.html}}. Generally, dust relevant for SNe is either graphite, with $\rho=2.26\,\mathrm{g}\,\mathrm{cm}^{-3}$, or silicate, with $\rho=3.30\,\mathrm{g}\,\mathrm{cm}^{-3}$. Therefore, there are two free parameters in the warm dust model: $M_\mathrm{d}$ and $T_\mathrm{d}$.

The total theoretical spectrum is then given by
\begin{eqnarray}
    f_\nu &=& F_{\nu,\mathrm{BB}}(R_{\mathrm{BB}}, T_{\mathrm{BB}}) + F_{\nu,\mathrm{d}}(M_\mathrm{d}, T_\mathrm{d}) \nonumber \\ 
    &=& \frac{\pi R_{\mathrm{BB}}^2}{d_\mathrm{L}^2}B_\nu(T_{\mathrm{BB}}) + \frac{3Q_\nu(a)}{4\rho a}\frac{M_\mathrm{d}}{d_\mathrm{L}^2}B_\nu(T_\mathrm{d}), 
\end{eqnarray}
from which we obtain the best-fit $R_{\mathrm{BB}}$, $T_{\mathrm{BB}}$, $M_\mathrm{d}$, and $T_\mathrm{d}$ through our fitting procedure. For this test, to ensure that our results are well-constrained, we only fit the model to the six epochs with $JHK$ measurements. Initially, we attempted two separate fits: one for silicate dust and the other for graphite dust. However, the fits for silicate dust consistently produced $T_\mathrm{d}\sim1700\,\mathrm{K}$, well above the evaporation temperature of silicate dust of $\sim 1100-1500\,\mathrm{K}$. Therefore, we did not further consider the case of silicate dust and only used the $Q_\nu(a)$ and $\rho$ values for graphite dust.

\begin{figure*}
\epsscale{1.10}
\plotone{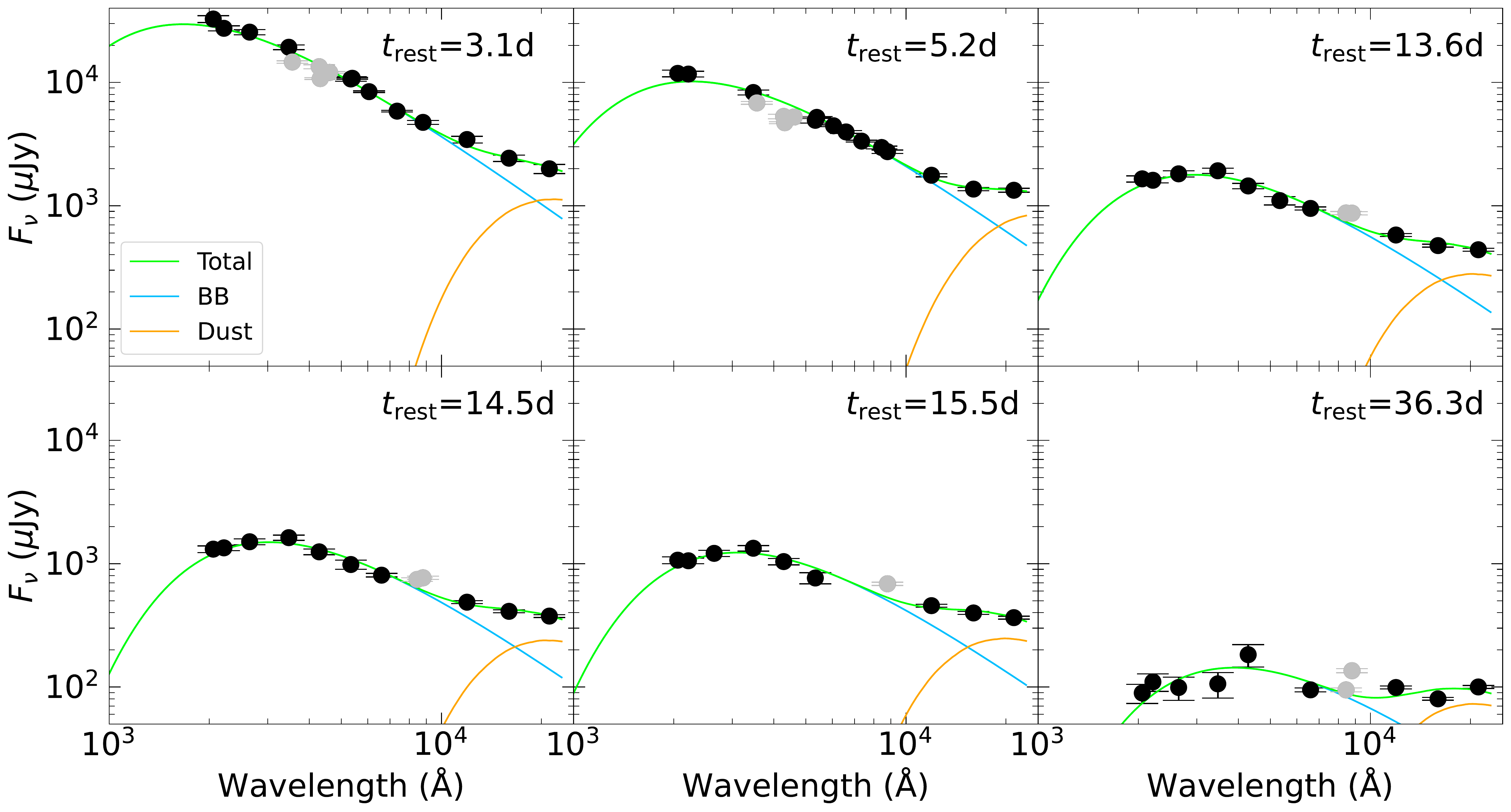}
\caption{The observed (dereddened) SEDs of AT\,2018cow (black circles) and the best-fit models (solid lines) derived from fitting the \textsc{blackbody + dust} model. The six chosen epochs all have JHK measurements that provide better constraints on the excess IR emission. Photometry excluded from the fit are shown in gray (see text for reasoning). The dust model generally provides a reasonable fit to the observed excess IR emission. More importantly, the UV-optical continuum from the is almost exclusively from the hot blackbody and unaffected by the dust model.
\label{fig:SEDDustfits}}
\end{figure*}

The observed SEDs with $JHK$ measurements, together with the best-fit models, are shown in Figure \ref{fig:SEDDustfits}. We found that even with the dust model, the shape and amplitude of the UV-optical continuum are almost exclusively dictated by the blackbody. On the other hand, the dust model does provide a reasonable fit to the observed excess IR emission.

We also derived the bolometric luminosities for the \textsc{blackbody + dust} model. The luminosity of the blackbody was still calculated using Equation \ref{eq:BBLum}. The luminosity of the dust was taken as 
\begin{equation}
L_{\mathrm{d}} = \frac{3\pi M_\mathrm{d}}{\rho a} \int_0^{\infty} Q_\nu B_\nu(T_\mathrm{d}) d\nu.
\end{equation}
The best-fit parameters, bolometric luminosities, and the reduced chi-square $\chi^2/d.o.f$ from fitting the \textsc{blackbody + dust} model are given in Table \ref{tab:testmodel}. The derived dust temperature ($T_{\mathrm{d}} \sim 1100-1400\,\mathrm{K}$) is below the evaporation temperature of graphite ($\sim 1900\,\mathrm{K}$). The derived mass ($M_\mathrm{d}\sim 10^{-4}-10^{-6}$) also seems reasonable for SNe \citep[e.g., see][for Type Ibn SNe]{Gan2021}. The $\chi^2/\mathrm{d.o.f}$ values are also similar to those from the \textsc{blackbody + power law} model.

Most importantly, the thermal properties (i.e., temperature and radius) derived from the \textsc{blackbody + dust} fits are very similar to those derived from the \textsc{blackbody + power law} fits. Therefore, we conclude that the thermal properties derived from the blackbody are robust and reliable for our interpretations in this study regardless of the description of the excess IR emission.

\begin{deluxetable*}{cccccccc}
\tablecaption{Properties of AT\,2018cow derived from the \textsc{blackbody + dust} model\label{tab:testmodel}}
\tablewidth{0pt}
\tablehead{
\colhead{$t\,(\mathrm{day})$} & \colhead{$R_{\mathrm{BB}}\,(R_\odot)$} & \colhead{$T_{\mathrm{BB}}\,(\mathrm{K})$} & \colhead{$M_{\mathrm{d}}\,(10^{-5}\,M_\odot)$} & \colhead{$T_{\mathrm{d}}\,(\mathrm{K})$} & \colhead{$L_{\mathrm{BB}}\,(10^{41}\,\mathrm{erg}\,\mathrm{s}^{-1})$} & \colhead{$L_{\mathrm{d}}\,(10^{41}\,\mathrm{erg}\,\mathrm{s}^{-1})$} & \colhead{$\chi^2/\mathrm{d.o.f}$}
}
\startdata
3.09  & $11139.47^{+240.73}_{-251.08}$ & $30513.00^{+911.27}_{-823.90}$ & $13.99^{+13.30}_{-5.60}$ & $1342.54^{+123.81}_{-132.48}$ & $3707.85^{+286.10}_{-243.24}$ & $8.42^{+1.05}_{-0.96}$ & 5.91 \\
5.20  & $10278.66^{+201.04}_{-208.99}$ & $22545.53^{+545.90}_{-510.42}$ & $23.06^{+10.82}_{-7.40}$ & $1147.50^{+71.62}_{-64.67}$   & $940.61^{+55.02}_{-48.99}$    & $5.31^{+0.32}_{-0.30}$ & 4.61 \\
13.56 & $6436.58^{+178.68}_{-182.62}$  & $17228.09^{+362.22}_{-339.02}$ & $2.62^{+0.67}_{-0.48}$   & $1420.19^{+56.79}_{-59.87}$   & $125.94^{+4.27}_{-4.17}$      & $2.14^{+0.12}_{-0.11}$ & 4.84 \\
14.53 & $6090.10^{+181.88}_{-183.84}$  & $16852.33^{+357.73}_{-337.74}$ & $2.50^{+0.62}_{-0.46}$   & $1388.99^{+55.78}_{-55.98}$   & $103.23^{+3.24}_{-3.26}$      & $1.79^{+0.10}_{-0.10}$ & 2.62 \\
15.54 & $5773.85^{+320.69}_{-297.70}$  & $16370.74^{+521.14}_{-503.43}$ & $2.06^{+0.53}_{-0.39}$   & $1454.01^{+65.78}_{-68.66}$   & $82.60^{+2.75}_{-2.60}$       & $1.93^{+0.16}_{-0.15}$ & 3.79 \\
36.27 & $2758.71^{+153.19}_{-151.79}$  & $13083.00^{+568.79}_{-499.48}$ & $0.70^{+0.13}_{-0.11}$   & $1411.00^{+46.70}_{-46.02}$   & $7.69^{+0.59}_{-0.50}$        & $0.55^{+0.03}_{-0.02}$ & 25.89  
\enddata
\tablecomments{Errors given here are statistical errors from the 15.9th and 84.1th percentile}. The degrees-of-freedom ($d.o.f$) is taken to be the number of data points minus the number of free model parameters.
\end{deluxetable*}

\clearpage

\bibliography{ref}{}
\bibliographystyle{aasjournal}

\end{document}